\documentclass[fleqn,usenatbib]{mnras}

\usepackage{newtxtext,newtxmath}
\usepackage[T1]{fontenc}

\DeclareRobustCommand{\VAN}[3]{#2}
\let\VANthebibliography\thebibliography
\def\thebibliography{\DeclareRobustCommand{\VAN}[3]{##3}\VANthebibliography}

\usepackage{graphicx}
\usepackage{amsmath}
\usepackage{caption}
\usepackage{subcaption}
\usepackage{tabularray}
\UseTblrLibrary{booktabs}
\usepackage{siunitx}
\usepackage{url}

\usepackage{color}
\definecolor{darkred}{RGB}{139, 0, 0}

\DeclareCaptionLabelFormat{figandsub}{\textbf{Figure \thefigure\alph{subfigure}.}}
\DefTblrTemplate{caption-tag}{default}{Table~\thetable}

\title[Are High-EW LAEs good analogs for the earliest galaxies?]{Are High-EW LAEs good analogs for the earliest galaxies? \\ A study on the ionizing conditions and gas properties of LAEs at $2\lesssim z \lesssim 7$}

\author[Lena Pless et al.]
{Lena Pless,$^{1}$\thanks{E-mail: lpless@swin.edu.au}
Themiya Nanayakkara,$^{1}$
Karl Glazebrook,$^{1}$
Michael Maseda,$^{2}$
Jarle Brinchmann,$^{3}$
\newauthor{and Tanya Urrutia$^{4}$}
\\ \\
$^{1}$Centre for Astrophysics and Supercomputing, Swinburne University of Technology, PO Box 218, Hawthorn, VIC 3122, Australia\\
$^{2}$Department of Astronomy, University of Wisconsin-Madison, 475 N. Charter St., Madison, WI 53706, USA\\
$^{3}$European Southern Observatory, Karl-Schwarzschild-Str. 2, D-85748 Garching, Germany\\
$^{4}$Leibniz-Institut für Astrophysik Potsdam (AIP), An der Sternwarte 16, D-14482 Potsdam, Germany
}

\date{Accepted XXX. Received YYY; in original form ZZZ}

\pubyear{\the\year{}}

\begin{document}
\label{firstpage}
\pagerange{\pageref{firstpage}--\pageref{lastpage}}
\maketitle

\begin{abstract}
Lyman-$\alpha$ emitters (LAEs) are frequently proposed as analogs of the low-mass, metal-poor galaxies thought to dominate cosmic reionization. In particular, galaxies with large Ly$\alpha$ equivalent widths (EWs) are expected to have not experienced significant chemical enrichment. We test this hypothesis using a sample of 88 LAEs at $2 \lesssim z \lesssim 7$, selected from MUSE observations undetected in deep HST imaging and followed up with JWST/NIRSpec spectroscopy. Despite the unprecedented sensitivity of JWST, only 46 of the LAEs exhibit secure counterparts in deep NIRCam imaging reaching typical 3$\sigma$ depths of $m_\text{AB}\sim30.3$ mag, highlighting the intrinsically faint nature of the population. Gas-phase metallicities, ionization parameters, and hydrogen densities are inferred from optical emission-line fitting with BPASS+cloudy photoionization models, and stellar masses from NIRCam SED fitting. The final sample with secure stellar masses and metallicities encompasses only 16 LAEs, occupying regions of standard emission-line diagnostic diagrams consistent with highly ionized, low-metallicity star-forming systems. Although the R3 diagnostic appears to decrease with increasing Ly$\alpha$ EW, we find no compelling evidence for a monotonic relation between Ly$\alpha$ EW and gas-phase metallicity when metallicities are inferred through full emission-line fitting.  However, galaxies with stronger Ly$\alpha$ emission tend to lie below the mass-metallicity relation at fixed stellar mass, suggesting they are preferentially less chemically evolved than typical galaxies of the same mass. We conclude that high-EW LAEs provide useful, though imperfect, reionization analogs, and that combining Ly$\alpha$ EW with stellar mass may more effectively identify the most chemically primitive galaxies.
\end{abstract}


%%%%%%%%%%%%%%%%%%%%%%%%%%%%%%%%%%%%%%%%%%%%%%%%%%

%%%%%%%%%%%%%%%%% BODY OF PAPER %%%%%%%%%%%%%%%%%%

\section{Introduction}
    A key prevailing question in extragalactic astronomy is understanding how the Universe underwent its last major phase transition: reionization. The onset of reionization is associated with the formation of the first stars and galaxies, and therefore fundamentally related to our understanding of galaxy formation and evolution \citep{Dijkstra_2014}. 
    Fully characterizing the sources responsible for reionization requires the observation of primordial galaxies. 
    Deep imaging obtained with the Hubble Space Telescope (HST) has allowed us to study in detail the post-reionization Universe, abundant with star-forming galaxies and largely transparent to light by a redshift of $z=5.5$, approximately one billion years after the Big Bang \citep{stark_2016}. 
    The launch of the James Webb Space Telescope (JWST) has immensely facilitated the discovery of galaxies past the end of reionization, out to redshifts $z\gtrsim10$ (see e.g. \cite{Finkelstein_2023, Labbe_2023, Mason_2023, Carniani_2024, Scholtz_2024}). 
    This extremely high redshift galaxy population exhibits large masses, large star-formation rates (SFRs), and is unexpectedly UV bright, with some galaxies showing high ionization emission lines, often with anomalously high equivalent widths (EWs) and/or unexpected line ratios such as GN-z11 \citep{Bunker_2023}, exhibiting strong emission in $\text{NIII}]\lambda1750$ and $\text{CIII}]\lambda\lambda1907,1909$ as well as super-solar N/O ratios, which are seemingly at odds with the young age of the Universe at a redshift of $z\sim11$, given the timescales of nitrogen enrichment \citep{Cameron_2023}. 
    Emission lines originating from $\text{HeII}$ (54.4 eV), $\text{CIV}$ (49.9 eV), or $\text{CIII}]$ (22.4 eV) require a significant amount of energy to be raised into an excited state, indicating extreme conditions in the gas, such as high temperatures or highly energetic irradiation. 
    Stellar population synthesis models tuned to low-redshift galaxies struggle to explain these extreme conditions and peculiar chemical enrichment patterns progressively uncovered in the JWST era. Whether such massive, UV-bright galaxies are the dominant drivers of reionization \citep{Naidu_2020a}, or whether reionization is instead sustained by a largely undetected population of lower-mass, faint galaxies \citep{Atek_2024}, remains an open and actively debated question. \\ \\ 
    Despite these recent observational advances into the high-redshift Universe, these low-mass galaxies still largely evade detection. While gravitational lensing has enabled the discovery of a small number of intrinsically faint, low-mass galaxies, a statistically representative sample remains beyond current observational capabilities. Identifying sources at redshifts $z>9$ typically requires galaxies to have already built up a significant amount of stellar mass and to be $ M_\text {UV} $- bright, leaving the youngest and lowest-mass galaxies largely undetected.
    It is thus still beneficial to look to analogous low-metallicity galaxies with low mass but highly ionizing properties at intermediate redshifts ($2<z<6$), analyzing their stellar and gas properties to gauge the contribution of these early, extreme sources to the reionization process. \\ \\
    The characteristics of the hot interstellar medium (ISM) in these early galaxy analogs can be pinpointed utilizing rest-frame optical emission lines \citep{nakajima_2023}. 
    Some emission lines and line ratios depend purely on the temperature of the gas, such as the $[\text{OIII}]\lambda4363/[\text{OIII}]\lambda5007$ ratio, while others probe the density of the medium, such as the $[\text{SII}]\lambda6716/[\text{SII}]\lambda6731$ ratio \citep{Stasinska_2009}. 
    Combining emission lines of different species permits the constraint of further properties of the ISM; for example, the BPT diagram (for Baldwin, Phillips, and Terlevich \cite{BPT_1981}) allows for the characterization of the hardness of the ionizing radiation. 
    Further diagnostic diagrams, such as the $[\text{OIII}]\lambda5007/\text{H}\beta$ vs $[\text{SII}]\lambda\lambda6716,6731/\text{H}\alpha$ \citep{veilleux_1987} diagram, help to gauge the ionization state, density, and metallicity of objects.
    With the launch of JWST, we now have an unprecedented view of the high-redshift universe, extending the range of redshifts at which rest-frame optical emission lines are observable beyond $z\approx2-3$, the limit for ground-based instruments. Nevertheless, the optical emission lines in the highest redshift sources remain elusive; another reason why it remains paramount to identify and study analogous, highly ionizing galaxies at intermediate redshifts.\\ \\
    A good candidate for intermediate redshift analogs to primordial galaxies is Ly$\alpha$ emitters (LAEs), as they have been determined to be highly star-forming, low-metallicity, and low-mass objects (see e.g. \citet{maseda_2018, maseda_2023, Roy_2023}. A galaxy is typically considered a LAE if the rest-frame EW is $\text{EW}_0\gtrsim20$ \AA\ \citep{Ouchi_2020}. Central Ly$\alpha$ emission occurs mainly due to ionization by young, hot, and massive stars or the presence of an AGN. As Ly$\alpha$ originates from a transition into the ground state, it may undergo resonant scattering. Thus, Ly$\alpha$ emission indirectly traces the distribution and dynamics of $\text{HI}$ gas in galaxies, allowing us to probe the neutral hydrogen gas in and around high-redshift objects, which is generally hard to investigate otherwise \citep{Wisotzki_2018}. Additionally, some recent studies have shown that strong Ly$\alpha$ emission traces galaxies with high ionizing photon escape fraction $\text{f}_\text{esc}^\text{LyC}$, which is necessary for galaxies to ionize their surrounding IGM during the EoR \citep{Begley_2024}.
    Theory predicts that high equivalent width LAEs ($\text{EW}_{0,\text{ Ly}\alpha}\gtrsim200$ \AA) are galactic systems that have yet to experience significant chemical enrichment, as Ly$\alpha$ emission is boosted in the surrounding gas \citep{raiter_2010, zackrisson_2011}, analogous to primordial galaxies. Recent observational studies have shown evidence that the Ly$\alpha$ EW of galaxies may indeed be used as a selection criterion for extremely low metallicity systems \citep{maseda_2023}. \\ \\
    Observations of nearby star-forming regions suggest that binary and higher multiplicity stellar systems are the norm even for low-mass stars \citep{Duchene_2013}, and latest findings show that the presence of massive stars in close binaries is ubiquitous even at low metallicity \citep{Sana_binaries_2025}. 
    Therefore, the effect of stellar multiplicity is important when modeling the evolution and properties of young stellar populations \citep{eldridge_2017}, and may explain the excess of highly ionizing radiation observed in many star-forming systems. The inclusion of binary evolution can lead to older stellar populations appearing more blue due to the contribution of interacting binaries \citep{eldridge_2009}, indicating higher amounts of ionizing radiation at later times. 
    While observations cannot yet be perfectly matched, including the effects of binary evolution, like the stellar population models \verb|BPASS|\citep{eldridge_2017}, appears to yield models that more accurately represent reality. Combining these models with photoionization codes such as \verb|cloudy| \citep{chatzikos_2023} assuming a range of abundance patterns (such as varying C/O and N/O ratios, which have been increasingly identified at high redshift \citep{Romano_2022, Isobe_2023}) may help to illuminate the nature of the gas properties in highly ionizing intermediate-redshift galaxies and thus primordial galaxies in the EoR. \\ \\
    In this work, we test whether high-EW LAEs satisfy the physical criteria expected of analogs to the low-mass galaxies thought to dominate reionization: exhibiting low stellar masses, low metallicities, and extreme ionization conditions. We analyze the properties of 88 galaxies that exhibit Ly$\alpha$ emission in the MUSE HUDF \citep{Bacon_2023} or MUSE Wide survey \citep{Herenz_2017, Urrutia_2019}, and determine their Ly$\alpha$ $\text{EW}_0$ utilizing NIRCam F090W observations to estimate the continuum emission as described in section \ref{sec:Data}. 
    Using NIRSpec MSA observations, optical emission line fluxes were measured to constrain ISM properties such as the metallicity, density, and ionization parameter (section \ref{sec:ISM_prop}) by fitting to an emission line diagnostic grid produced using \verb|BPASS| v2.2.1 and \verb|cloudy| v23.01 as described in section \ref{sec:Grid}. The results and discussion are presented in section \ref{sec:ISM_prop} and \ref{sec:res_disc}, followed by a summary and main conclusions in section \ref{sec:conclusions}. \\ \\
    Throughout this article, we adopt a solar abundance of $\text{Z}_\odot=0.01524$ corresponding to $12+\log(\text{O/H})=8.69$ and a solar C/O ratio of $(\text{C/O})_\odot=0.44$ \citep{gutkin_2016}. Where needed, we adopt the \textit{Planck} 2018 flat $\Lambda$CDM cosmology \citep{PlanckCollaboration_2020} with $H_0=67.7\text{ km s}^{-1}\text{ Mpc}^{-1}$, $\Omega_\text{M}=0.31$, and $\Omega_\Lambda=0.69$.
\section{Data}\label{sec:Data}
\begin{figure}
        \centering
        \includegraphics[width=\linewidth]{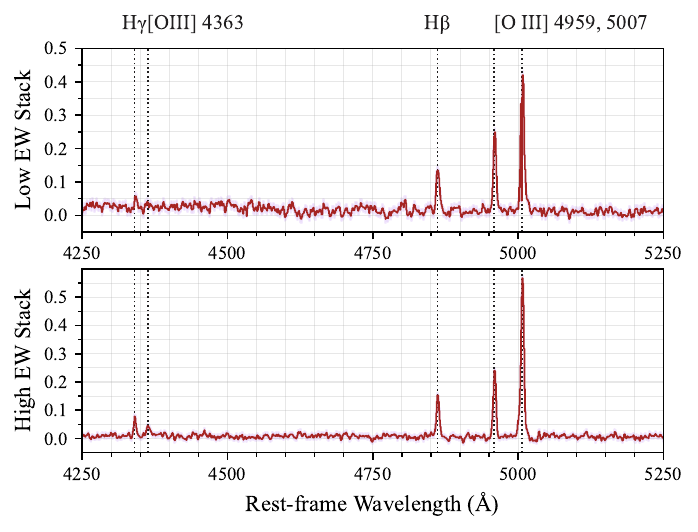}
        \caption{Low and high EW stacked spectra in the range of the {[OIII]$_{4959, 5007}$} lines, {[OIII]$_{4363}$} auroral line, and H$\beta$, H$\gamma$ emission lines}
        \label{fig:stacks_example}
    \end{figure}

    The sample of 88 LAEs that are the focus of this work was observed with JWST/NIRSpec’s Micro Shutter Array (MSA) in the G235M/F190LP and G395M/F270LP configuration as presented in \cite{maseda_2023} (JWST Program ID 1671, PI: Maseda).  
    Primary targets were selected from MUSE HUDF \citep{Bacon_2023} and MUSE WIDE \citep{Herenz_2017, Urrutia_2019} to show significant Ly$\alpha$ emission (SNR > 5) while being UV-continuum faint, i.e., being extremely faint or undetected in available HST imaging,  suggesting a very faint continuum ($m_\text{AB} > 30$ mag).
    Additionally, ELGs selected from 3D-HST grism data \citep{Momcheva_2016} with low continuum emission, akin to the LAEs, are included in the final sample. 
    Of this sample of 112 highly ionizing galaxies, 88 show Ly$\alpha$ emission as observed with MUSE, which are the focus of this work.
    %\footnote{With the second data release of MUSE WIDE, Ly$\alpha$ has been detected in four targets selected from 3D-HST, extending the sample to 88 LAEs.}
    The NIRSpec MSA spectra were re-reduced using the level 1 and level 2 of the \verb|jwst| version 1.14.0 pipeline and the \verb|jwst_1214.pmap| context \citep{Bushouse_JWST_Calibration_Pipeline_2024} and \verb|msaexp| version 0.8.4 \citep{Brammer_msaexp_NIRSpec_analyis_2022} to drizzle and combine all exposures. \\ \\
    To investigate average spectral properties as a function of Ly$\alpha$ EW within the sample, all spectra were stacked to increase the SNR in faint lines. 
    The stacked spectra were constructed by combining the individual spectra of galaxies grouped by Ly$\alpha$ equivalent width. 
    Galaxies were divided into high- ($\text{EW}_{0,\text{ Ly}\alpha}\geq90$ \AA) and low-EW ($\text{EW}_{0,\text{ Ly}\alpha}<90$ \AA) subsamples as in \citet{maseda_2023}. 
    For each group, spectra were shifted to a common rest-frame wavelength grid using each source's systemic redshift, and flux values at each wavelength were averaged across all galaxies in the respective subsample. 
    Prominent emission lines in the stacked spectra are shown in Figure \ref{fig:stacks_example}.
    \noindent \\ \\ Emission lines in the spectra were fit with the Gaussian line fitting code \verb|jwst-msafit| \citep{de_graaf_msafit_2024}, which takes slit losses and line spread functions for any user-specified morphology into account.
    Lines (see Tab. \ref{tab:emission_lines}) were fitted on a combined spectrum of the G235M/F190lp and G395M/F270lp configuration, where fluxes are scaled by the overlap between the two obtained spectra.
    In cases with no overlapping wavelength coverage, lines were fitted separately at a common redshift solution. 
    \begin{table}
        \centering
        \begin{tabular}{p{0.7\linewidth}|p{0.2\linewidth}}
            \textbf{Stage} & \textbf{\# of targets} \\ \hline 
             Initial set of LAEs \& ELGs & 112 \\
             LAEs & 88 \\
             LAEs with EW measurements & 82 \\
             LAEs with NIRCam counterparts & 46 \\
             LAEs with reliable masses & 36\\
             LAEs $\geq3$ emission lines in NIRSpec & 21 \\
             LAEs with reliable metallicities & 20 \\
             LAEs with reliable metallicities \& masses & 16
        \end{tabular}
        \label{tab:sample}
        \caption{Number of targets given the different observational and analytical constraints}
    \end{table}
\subsection{Slit Positioning}\label{sec:slit_positioning}
\begin{figure}
        \centering
        \includegraphics[width=\linewidth]{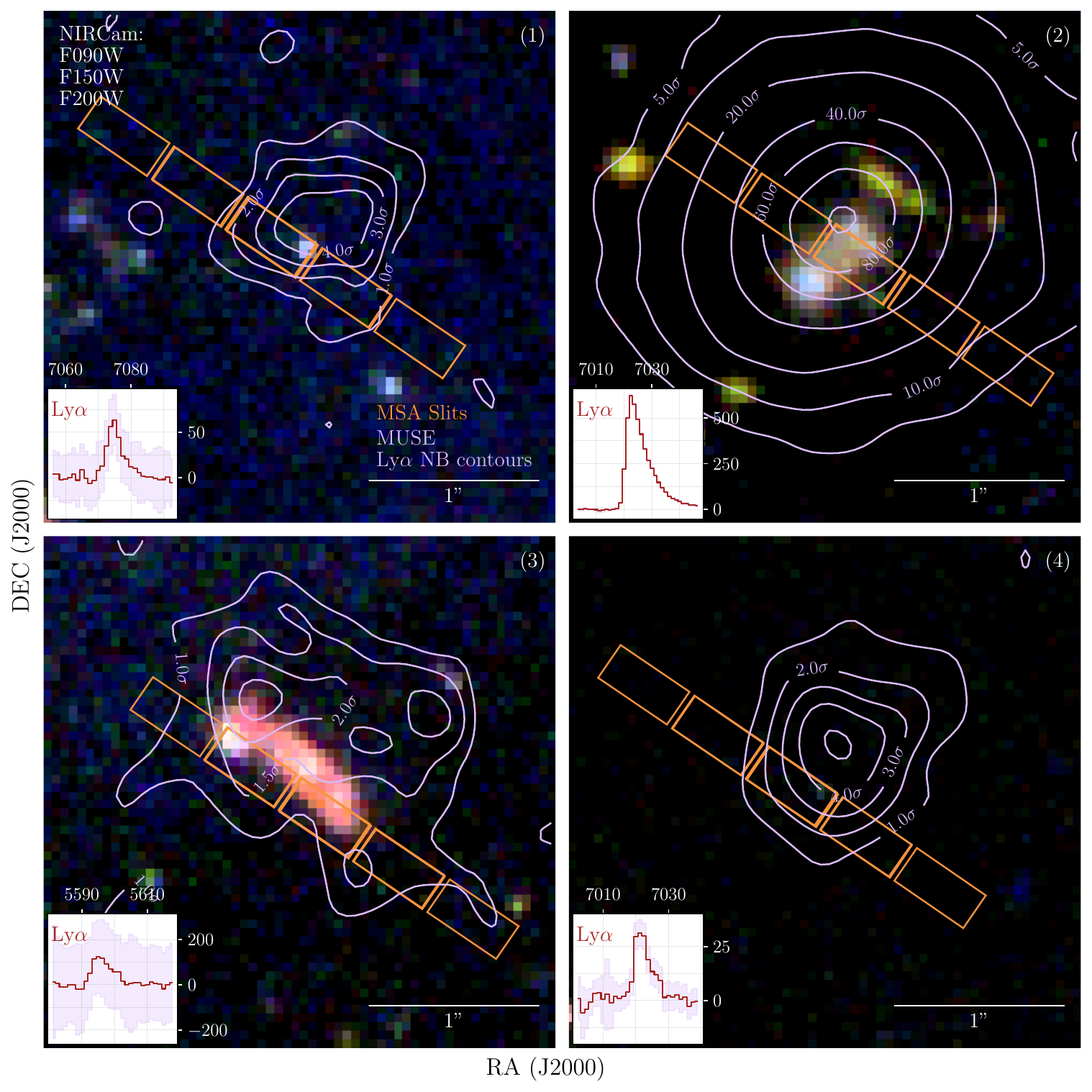}
        \caption{Example of LAEs as seen in NIRCam (RGB F090W, F150W, F200W) with MUSE Ly$\alpha$ NB contours and integrated Ly$\alpha$ profiles shown in the insets \\ (1) strong Ly$\alpha$ emission with an associated host (point source)\\ (2) strong Ly$\alpha$ emission with an associated host (extended) \\ (3) weak Ly$\alpha$ emission with associated host (extended, offset) \\ (4) weak Ly$\alpha$ emission and no associated host}
        \label{fig:example_LAEs}
    \end{figure}

\begin{figure}
        \centering
        \includegraphics[width=\linewidth]{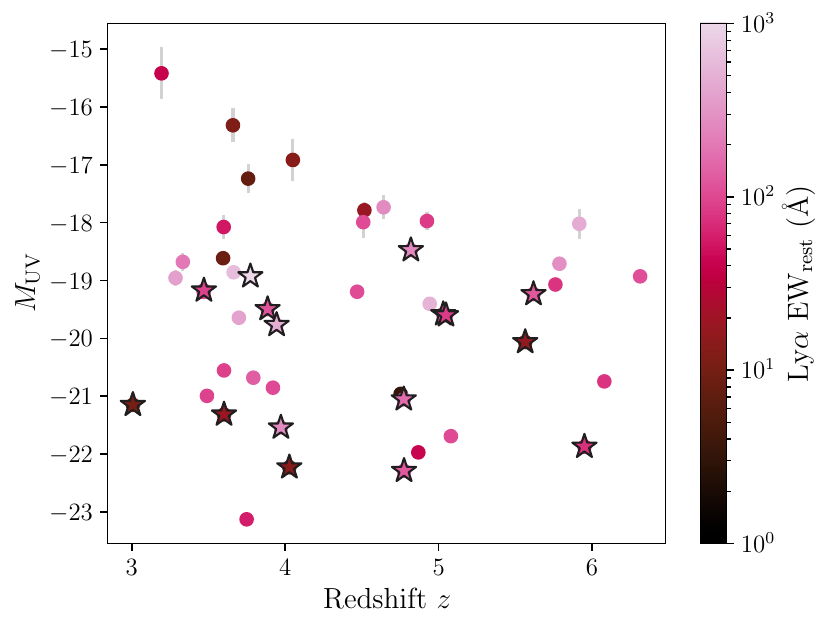}
        \caption{Absolute Magnitude $M_\text{UV}$ versus redshift for the $N=46$ LAEs with counterparts colored by their determined EW. The final sample of LAEs passing all quality cuts ($N=16$) is marked with stars.}
        \label{fig:muv_vs_z}
    \end{figure}

    Around a third of the obtained spectra appear to be noise-dominated, with no significant emission lines detected in the rest-frame optical. 
    This may partially be due to the frequently observed offset between Ly$\alpha$ emission and the center of galaxies (see e.g. \citet{Hoag_2019}). 
    As targets are selected to be continuum-faint, and NIRCam imaging was not available at the time of data collection, this known offset leads to discrepancies between the galaxy and slit positions, which are centered on the Ly$\alpha$ emission as observed with MUSE. 
    Out of the 88 LAEs, clear association between the Ly$\alpha$ emission and a host using recent NIRCam imaging surveys (FRESCO, JADES, and JEMS) in at least one of the filters covering GOODS-S, both in spatial and redshift space, can be made for 46 of these sources, examples of which are shown in Fig. \ref{fig:example_LAEs}. For a clear association, we require the Ly$\alpha$ emission and a host in imaging to be within a radius of 1 arcsecond and at a comparable redshift with $\Delta z\leq0.01$ in the photometric catalogs available from FRESCO, JADES, and JEMS. Due to the redshift constraint, no association may be possible if there are several faint sources with no redshift information in the photometric catalogs. Thus, the 42 targets without clear associations are made up of targets undetected in NIRCam, as well as targets for which no redshift is available to confirm association. Examples of these sources are shown in Appendix \ref{fig:non-detections}. \\ \\
    However, only 25 targets show significant optical line detections with $\text{SNR}\geq3$ in at least 3 emission lines in the NIRSpec spectra. This significant decrease in the LAE sample is in part due to the faintness of the targets, but can also be partially attributed to adverse slit positioning, with 11 LAE counterparts not being covered by the slit footprints of the survey. 
    Of the 25 LAEs with significant emission, we obtain well-constrained metallicity measurements and stellar mass estimates for 16 targets. This subsample spans the full range of Ly$\alpha$ EWs observed in the parent LAE sample. Figure \ref{fig:ew_hist} shows that it reproduces the overall shape of the EW distribution, suggesting that it is broadly representative of the full LAE sample, though the low EW bin ($<90$ \AA) is more sparsely sampled than the high EW bin ($\geq90$ \AA), and no targets exhibiting equivalent widths between $20$ \AA and $50$ \AA are included in the final subsample. Fig. \ref{fig:muv_vs_z} shows the sample's UV absolute magnitude versus redshift, with the final sample passing all quality cuts (stars) spanning almost the full redshift range $3\lesssim z\lesssim6$. However, the final sample is limited to only the brighter targets with $M_\text{UV}\lesssim-18$ mag, with the faintest objects dropping out due to slit positioning, an unconstrained SED, or insufficient emission lines at $\text{SNR}\geq3$.  
\subsection{Ly$\alpha$ EW Measurement}
\begin{figure}
        \centering
        \includegraphics[width=\linewidth]{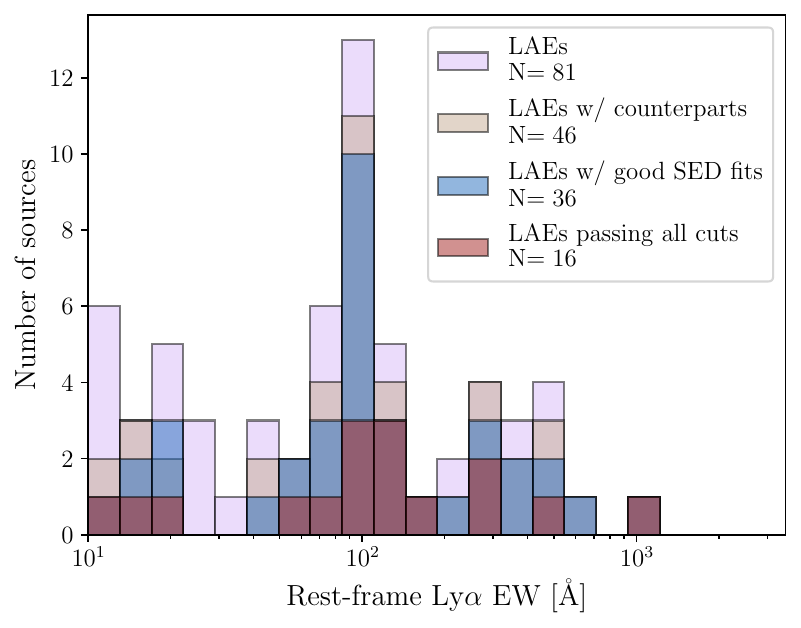}
        \caption{Histogram of rest-frame Ly$\alpha$ equivalent widths for the full LAE sample with EW measurements (mauve; $N=81$) and the consecutive subsamples with counterparts (khaki; $N=46$), good SED fits (blue; $N=36$) and reliable metallicity measurements (red; $N=16$). The reliable metallicity subsample covers the full range of equivalent widths present in the LAE sample and exhibits a similar overall distribution.}
        \label{fig:ew_hist}
    \end{figure}

    Previous measurements of the rest-frame Ly$\alpha$ equivalent widths relied on continuum estimates derived either from the extracted MUSE spectra \citep{maseda_2023} or from HST imaging \citep{Kerutt_2022}. Since targets were specifically selected to be continuum-faint in HST and deep JWST/NIRCam observations are now available, we remeasure the rest-frame equivalent widths, $\text{EW}_{0,\mathrm{Ly}\alpha}$, for 81 LAEs of the 88 sample using F090W imaging data to estimate the rest-frame UV continuum flux density. One target exhibited an anomalous spectrum; 6 more could not be included in the analysis as they were cut from the MUSE WIDE DR2, which applies a more stringent SNR cut ($\text{SNR}\geq5$) than DR1 ($\text{SNR}\geq3$). For N=19 targets with redshift $z\gtrsim5.54$, Ly$\alpha$ falls into F090W, so F115W was used to estimate the continuum flux instead.\\ \\
    To determine the size of the emission region, \verb|photutils|'s \verb|detect_sources| was utilized on cutouts of 3x3" around the position of the Ly$\alpha$ emission. 
    If the target has a clear counterpart within a 1.5" radius, the size of the emission region was determined using the image with the highest SNR, thereby avoiding the exclusion of faint emission at the edges emission at the edge of the counterparts. 
    Otherwise, a fixed circular aperture with a 3-pixel radius centered on the Ly$\alpha$ position was used to determine an upper limit on the continuum flux, as sources are expected to be compact.
    The measured continuum flux was extrapolated to the Ly$\alpha$ line position using the continuum slope $\beta$ of the spectrum, which correlates the flux density $f$ at a certain wavelength $\lambda$ via $f_\lambda\propto\lambda^\beta$.
    Where at least two NIRCam bands cover the rest-frame UV continuum (here defined as $1250\leq\lambda\leq3000$ \AA), we fit a power law to the available photometry to determine the slope of the continuum.
    The mean UV continuum slope recovered is $\beta=-1.93\pm0.37$, in agreement with previous measurements of UV continuum slopes in the MUSE Deep and MUSE WIDE fields \citep{Kerutt_2022}. 
    For targets with insufficient filter coverage by NIRCam or non-detections, we utilized the mean recovered value, which is in concordance with the anticipated UV continuum slope for a moderately young and metal-poor stellar population $\left<\beta\right>\sim-2.0$ \citep{Cullen_2023}. For galaxies at $z\sim2$-$3$, F090W probes the rest-frame near-UV ($\sim2000$-$3000$ \AA), meaning the extrapolation to Ly$\alpha$ assumes the UV slope measured at these wavelengths is representative of the far-UV slope shortward of 2000 \AA. These two regimes can differ, as the far-UV and near-UV slopes are not necessarily continuous, potentially introducing a systematic uncertainty in the continuum estimate at the Ly$\alpha$ position for the lowest-redshift targets in the sample. We note this as a caveat but do not expect it to significantly affect the results, as the mean recovered $\beta$ is consistent with previous measurements and the anticipated slopes for young, metal-poor stellar populations \citep{Cullen_2023}. Additionally, the final sample of $N=16$ targets for which well-constrained metallicities are obtained spans redshift $3\lesssim z\lesssim6$ and is thus largely unaffected by this simplification.
    The rest-frame equivalent width was then estimated as the ratio of the flux in Ly$\alpha$ and the extrapolated UV continuum flux, adjusted for redshift. The distribution of recovered EW measurements is shown in Fig. \ref{fig:ew_hist}
\subsection{Stellar Masses}
    The SED fitting was performed using the Python tool $\verb|bagpipes|$ \citep{Carnall_2018}, assuming the \cite{Calzetti_2000} dust extinction law. As the targets are selected to be highly ionizing ELGs and most targets exhibit large equivalent widths in $\text{H}\alpha$ and $[\text{OIII}]$, the SED fitting is performed assuming a double-power law star formation history (SFH) to model a recent burst in star formation.  Additionally, the fitting is also performed with a delayed-exponential SFH and a bursty SFH to compare with the double-power law model and verify masses, as bursty SFHs may bias fitting results towards lower masses \citep{Sorba_2018}. \\ \\
    To fit the nebular emission, v2.2.1 of the $\verb|BPASS|$ \citep{Stanway_2018} stellar grids, employing a Salpeter IMF \citep{Salpeter_1955} and an upper mass threshold of $300M_\odot$, are utilized. Nebular emission line grids and continuum grids from those stellar models, produced with the photoionization code $\verb|cloudy|$ \citep{Ferland_2023}, implemented in $\verb|bagpipes|$, were used. 
    As sources are dominated by their nebular emission, spectroscopic data are fit alongside the photometry \citep{Carnall_2019}, since photometric bands may be dominated by flux from strong emission lines. 
    Because the sources are selected to be UV-faint, some targets are not significantly detected in the photometry, so fluxes are given as upper limits on the actual emission. SED fitting results are only put to use for sources with $\geq3\sigma$ detections in at least four photometric bands to ensure sufficient constraints for parameter estimation. 
\section{Emission Line Diagnostic Grid}\label{sec:Grid}
    To determine gas properties in star-forming galaxies, measured emission line ratios can be compared with predictions from numerical models. To accurately determine the gas and stellar properties of the sample of LAEs, the stellar population synthesis (SPS) models produced by the Binary Population and Spectral Synthesis (\verb|BPASS|) code \citep{eldridge_2017} are utilized. 
    The \verb|BPASS| v2.2.1 \citep{stanway_eldridge_2018} models not only include the effects of binary evolution but also span a large range of metallicities, down to $Z=0.00001$, equivalent to roughly $\sim0.05\%\text{ Z}_\odot$ and incorporate the stellar atmosphere models \verb|BASEL|  v3.1 \citep{Westera_2002}, supplemented with WR stellar atmosphere models \verb|PoWR| \citep{Hamann_2003, Sander_2015, Todt_2015} and \verb|WMBASIC| models based on the grid determined by \citet{Smith_2002} with extended metallicity coverage and higher surface gravity for O-type stars.
    \cite{Steidel_2016} find that \verb|BPASS| models v2.1 \citep{eldridge_2017} better reproduce the nebular emission in their sample of star-forming galaxies at $z\sim2.4$ in comparison to SPS models that do not include binary effects. \\ \\
    By combining the synthetic spectra of the $\verb|BPASS|$ code v2.2.1 \citep{stanway_eldridge_2018} and the photoionization code \verb|cloudy| v23.01 \citep{chatzikos_2023, gunasekera_2023}, a suite of emission line predictions was produced. 
    This new grid of photoionization models is constructed following the approach of \cite{gutkin_2016}, who used the \cite{bruzual_charlot_2003} SPS models, which do not include binary effects, and the most up-to-date stellar atmosphere models. \\ \\
    Previous work by \cite{xiao_2018} does present photoionization grids using $\verb|BPASS|$, however, unlike \cite{gutkin_2016} they do not consider the effects of dust, metal-to-dust depletion, and varying abundance patterns, particularly enhanced C/O and N/O ratios, which are prominent in high redshift galaxies according to recent JWST findings (see e.g. \cite{Isobe_2023}). \\ \\ 
    In the following, model spectra following a Salpeter IMF and an upper mass cut-off of $m_\text{up}=300\text{ M}_\odot$ have been chosen. 
    The $\verb|BPASS|$ binary stellar models range in age of their simple, coeval stellar population from 1 Myr to 100 Gyr in steps of $\log(\text{age yr}^{-1})=0.1$. 
    For the grid of model emission lines, the age range has been limited to 1-100 Myr, as young, hot stars are the primary source of ionizing radiation that drives emission lines in spectra of star-forming galaxies. \\ \\ 
    Using the approach in \citet{gutkin_2016}, the grid of emission line spectra is sampled across several parameters as outlined in Tab. \ref{tab:param_grid}: the metallicity of the gas $Z_\text{ISM}$ (assumed to be equivalent to the stellar metallicity); hydrogen density of the gas $\log(n_\text{H})$; strength of the incident ionizing radiation field as set by the zero-age ionization parameter at the Strömgren radius $\log(U_\text{S})$ (in the following $\log(U)$); the amount of metal depletion onto dust by setting a fixed dust-to-metal mass ratio characterized by the dust-to-metal mass ratio in the range $0\leq\xi_\text{d}\leq1$; the carbon-to-oxygen abundance ratio C/O (where $(\text{C/O})_\odot=0.44$); and the age of the stellar population (see Tab.\ref{tab:param_grid}). 
    \begin{table}
        \centering
        \begin{tabular}{p{0.2\linewidth}|p{0.7\linewidth}}
            \textbf{Parameter} & \textbf{Values} \\ \hline 
            $Z_\text{ISM}$ & \num{1.0e-4}, \num{2.0e-4}, \num{5.0e-4}, \\ 
            $\text{Z}_\odot=0.01524$& \num{1.0e-3}, \num{2.0e-3}, \num{4.0e-3}, \num{6.0e-3}, \\ 
            & \num{8.0e-3}, \num{1.0e-2}, \num{1.4e-2}, \num{1.7e-2}, \\
            & \num{2.0e-2}, \num{3.0e-2}, \num{4.0e-2} \\
            $\log(n_\text{H})$ & 1.0, 2.0, 3.0, 4.0 \\
            $\log(U)$ & -4.0, -3.5, -3.0, -2.5, -2.0, -1.5, -1.0 \\ 
            $\xi_\text{d}$ & 0.1, 0.3, 0.5 \\
            (C/O)/(C/O)$_\odot$ & 0.10, 0.14, 0.20, 0.27, 0.38, 0.52, 0.72, 1.00, 1.40 \\
            age yr$^{-1}$ & \num[print-exponent-implicit-plus]{1e+6} to \num[print-exponent-implicit-plus]{1e+8} in steps of $\log(\text{age yr}^{-1})=0.1$
        \end{tabular}
        \caption{Grid sampling of the parameters used for \texttt{cloudy} v23.01 following the setup presented in \citet{gutkin_2016}}
        \label{tab:param_grid}
    \end{table}
    The interstellar abundances and their corresponding depletion factors are also adopted from \cite{gutkin_2016}. Similarly, the nitrogen abundance is determined based on the oxygen abundance according to the following prescription described by \cite{groves_2004b}:
    \begin{equation}
        \text{N/H}\approx0.41\text{O/H}\left[10^{-1.6}+10^{2.33+\log(\text{O/H})}\right]
        \label{eq:n_abund}
    \end{equation}
    This results in nitrogen abundances ranging from local values $\log(\text{N/O})\approx-2.0$ to nitrogen enhancements similar to those observed at high redshift with JWST $\log(\text{N/O})\approx-1.0$ \citep{Zhu_2026}. 
    To preserve the set metallicity $Z_\text{ISM}$, the abundances of all elements heavier than $\text{He}$ are rescaled accordingly. 
    \verb|cloudy| v23.01 was run assuming a closed, spherical gas geometry with a covering factor of 1.0 around the stellar population starting at an inner radius of 1pc. Dust is included using the standard implemented \verb|ISM| prescription for dust grains. 
    \begin{figure*}
        \centering
        \includegraphics[width=\linewidth]{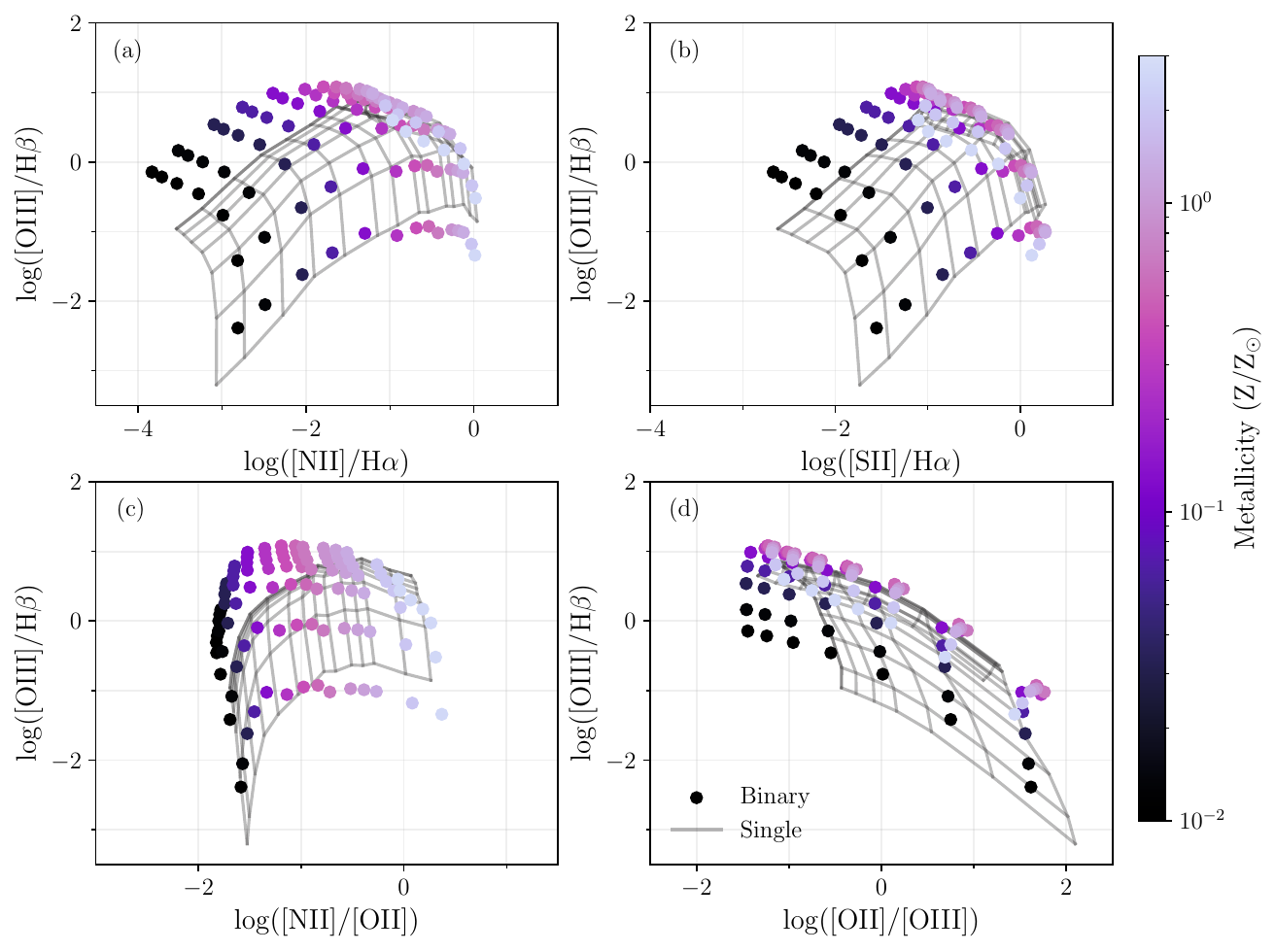}
        \caption{Optical emission line diagnostic plots based on \texttt{BPASS} v2.2.1 and \texttt{cloudy} v23.01 at fixed density $\log(n_\text{H})=3.0$. Additionally we show the same grid as lines for the single star models available in \texttt{BPASS} v2.2.1. While the overall shape of the grid stays consistent, the single star models do not reach as extreme line ratio values more commonly observed in the high redshift universe. With some of the LAEs analyzed here reaching $\log([\text{OIII}]/\text{H}\beta)\sim1.0$, single star models would struggle to determine a suitable metallicity solution.}
        \label{fig:CO_44_optical_bpt}
    \end{figure*}
    The resulting grid is shown in Fig.\ref{fig:CO_44_optical_bpt} for the most commonly used emission line diagnostic plots ([NII] BPT, [SII] BPT, [NII]/[OII] BPT and [OII] BPT) assuming a solar C/O ratio $(\text{C/O})_\odot=0.44$ at a hydrogen density of $\log(n_\text{H})=3.0$ for a constant SFH (i.e, integrating over all ages from 1 Myr to 100 Myr). \\ \\
    The resulting model grid shows trends similar to those in \cite{gutkin_2016}, with the grid presented here merely shifting it towards regions of stronger ionization, as expected given the observed differences between SPS models such as \verb|BC03| and \verb|BPASS|. There are, however, notable differences from the results obtained by \cite{xiao_2018}, particularly at higher metallicities.
    One of the main differences between the photoionization code setup by \cite{xiao_2018} and the grid presented here is the scaling of the nitrogen abundance according to Eq.\ref{eq:n_abund}. 
    This effect, however, is minimal and, unlike the observed discrepancy, occurs to the same extent across all metallicities. 
    Another possible reason for the discrepancy between the two grids could be that \cite{xiao_2018} do not include a dust prescription in their \verb|cloudy| models and thus do not account for metal depletion onto dust grains. 
    Indeed, when no metal-to-dust depletion is assumed, optical emission line ratios such as $[\text{NII}]/\text{H}\alpha$ and $[\text{OIII}]/\text{H}\beta$ decrease, as there is a larger number of metal coolants present, the electron temperature decreases, and thus cooling through collisionally excited optical lines is reduced as well \citep{gutkin_2016}. 
    This effect is stronger at higher metallicities and may contribute to the differences between the models presented here and those of \cite{xiao_2018}. 
\subsection{Bayesian Emission-Line Fitting}
    Physical parameters were inferred through Bayesian fitting of the observed emission-line fluxes to the BPASS photoionization model grid using a custom routine based on the code \verb|PIFit| (Brinchmann, private communication). The model grid spans gas-phase metallicity, ionization parameter, and hydrogen density over the parameter ranges described in Tab. \ref{tab:param_grid}. Stellar population emission line predictions were generated assuming a constant star-formation history and integrated uniformly over BPASS ages from 10 Myr to 100 Myr, such that stellar age was not treated as an independent free parameter during the fitting process. \\\\ For each source, all emission lines detailed in Tab. \ref{tab:emission_lines} covered by the observed spectral range were included in the fit.
    \begin{table}
        \centering
        \begin{tabular}{p{0.3\linewidth}|p{0.6\linewidth}}
            \textbf{Species} & \textbf{Lines included in fit}  \\ \hline
            H (Balmer series) & H$\alpha$, H$\beta$, H$\gamma$, H$\delta$ \\
            O & $[\text{OII}]_{3726, 3729}$, $[\text{OIII}]_{4363}$, $[\text{OIII}]_{4960, 5008}$ \\
            N & $[\text{NII}]_{6550, 6585}$ \\
            S & $[\text{SII}]_{6718, 6733}$ \\
            He & HeII$_{4686}$ \\
            Ne & $[\text{NeIII}]_{3869}$
        \end{tabular}
        \caption{Emission lines identified in the observational spectra. Grating coverage can result in only a subset of lines being fit in individual spectra.}
        \label{tab:emission_lines}
    \end{table}
    Rather than fitting line ratios, the observed absolute line fluxes were compared directly to the model predictions. As the BPASS grid provides line intensities relative to H$\beta$, a multiplicative normalization factor was introduced as a nuisance parameter to scale the model emission-line predictions to the observed flux level. This normalization absorbs differences in the overall luminosity scale while preserving the relative line strengths that constrain the physical conditions of the ionized gas. \\ \\ Posterior distributions were sampled using the nested-sampling algorithm \verb|MultiNest|. Flat priors were adopted for all fitted parameters. The likelihood function was assumed to be Gaussian and was computed as $-\frac{1}{2}\sum_i\left(\frac{A\cdot F_{i,\mathrm{mod}} - F_{i,\mathrm{obs}}}{\sigma_i}\right)^2$ where $F_{i,\mathrm{obs}}$ and $\sigma_i$ are the measured emission-line fluxes and associated uncertainties, $F_{i,\mathrm{mod}}$ is the model-predicted flux, and $A$ is the normalization factor. Consequently, all measured emission lines contribute to the fit, with low-significance measurements naturally receiving lower statistical weight through their larger uncertainties. \\ \\ The full BPASS photoionization grid includes additional parameters such as the dust-to-metal ratio and carbon-to-oxygen abundance ratio. However, the limited number of available emission-line constraints is insufficient to robustly constrain the full parameter space. To reduce degeneracies, these quantities were fixed to $\xi_{\rm d}=0.3$ and $\mathrm{C/O}=0.44$ (solar), respectively. \\ \\ Posterior samples and diagnostic corner plots were generated for each source. Throughout this work, the median of the marginalized posterior distribution is adopted as the fiducial parameter estimate, while the 16th and 84th percentiles are reported as the corresponding $1\sigma$ credible intervals.
\subsection{Validating Grid Predictions}
    \begin{figure}
        \centering
        \includegraphics[width=\linewidth]{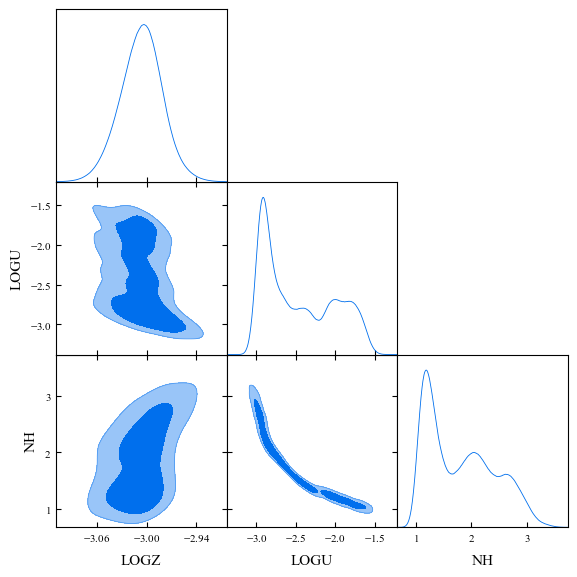}
        \caption{Cornerplot for a sample fit of emulated observational data. Posteriors for $\log(U)$ and $\log(n_\text{H})$ span large ranges as they are not well constrained using the set of emission lines present in the observations (see Tab. \ref{tab:emission_lines}). The posterior distribution for $\log(U)$ and $\log(n_\text{H})$ is characteristic for most fits, and solutions appear to be degenerate.}
        \label{fig:cornerplot}
    \end{figure}
    \begin{figure*}
        \centering
        \includegraphics[width=\textwidth]{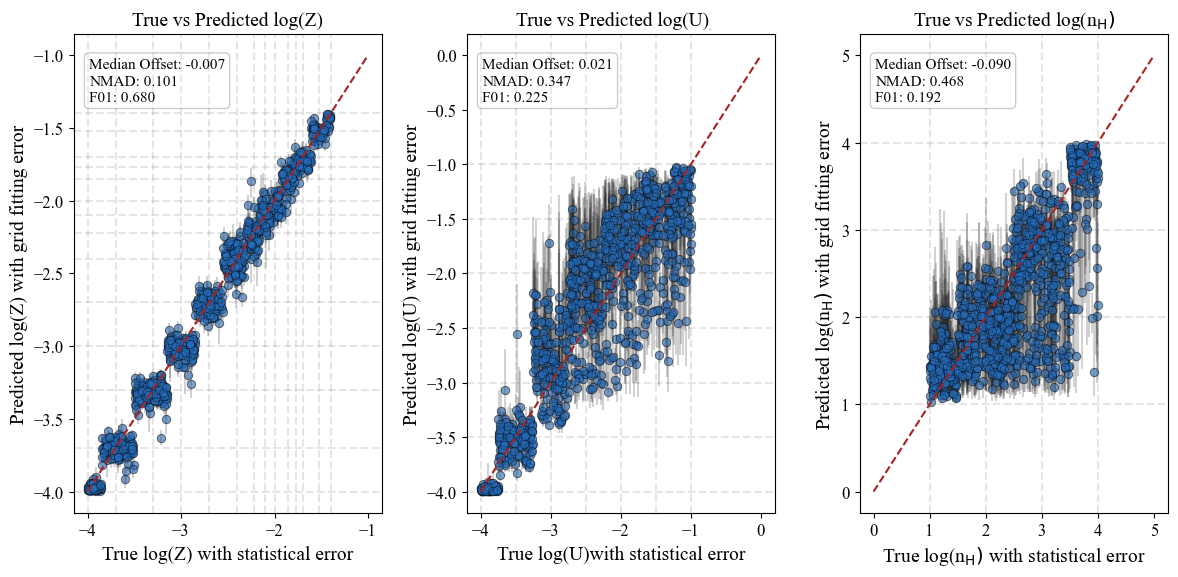}
        \caption{Results of $N=1000$ Monte-Carlo realizations of fitting synthetic emission line fluxes with errors comparable to the highest quality observed spectra to the grid. Metallicities ($\log(Z_\text{ISM})$) seem to be recovered well, while the ionization parameter ($\log(U)$) and density ($\log(n_\text{H})$) are not well constrained.}
        \label{fig:grid_mc_1000}
    \end{figure*}
    \begin{figure}
        \centering
        \includegraphics[width=\linewidth]{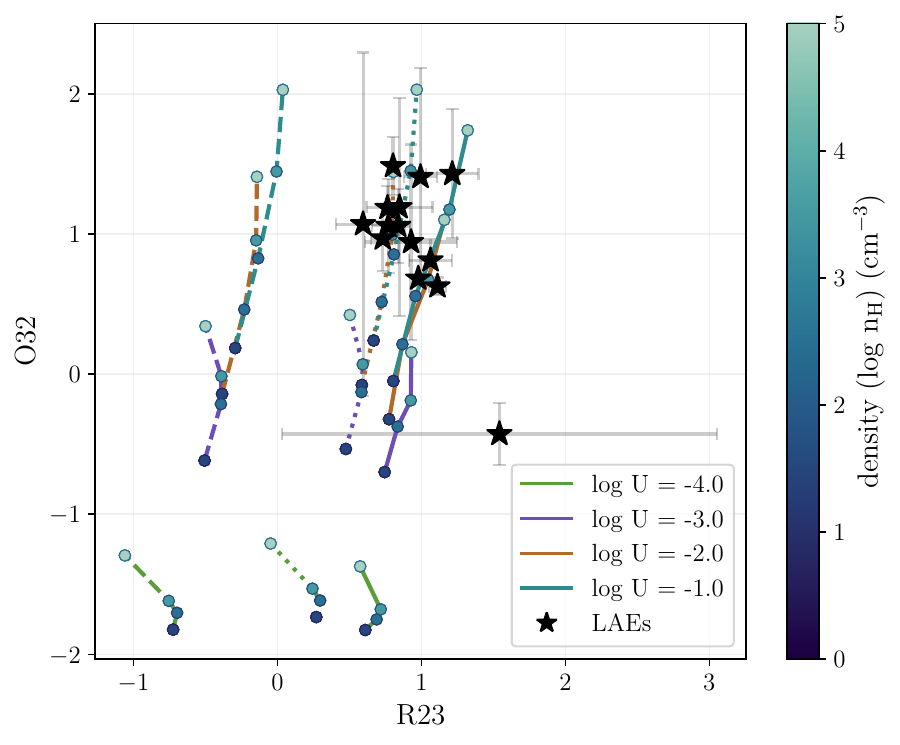}
        \caption{O32 as a function of R23 for the BPASS photoionization model grid. Tracks are shown for different ionization parameters ($\log(U) = -4.0$, $-3.0$, $-2.0$, and $-1.0$) at different metallicities ($1\%\text{ Z}_\odot$, $10\%\text{ Z}_\odot$ and Z$_\odot$). The color scale indicates hydrogen number density ($\log(n_{\mathrm{H}})$ in cm$^{-3}$). LAEs are shown as black stars, generally residing in the high-excitation region of this line-ratio space, though given the errors on measurements and the overlap of model predictions, it is difficult to recover unique solutions for $\log(U)$ and $\log(n_{\mathrm{H}})$.}
        \label{fig:r23_o32_grid_degeneracy}
    \end{figure}  
    To assess the performance of the grid-fitting procedure, a forward-model parameter-recovery test is performed using $N=1000$ Monte Carlo realizations.
    First, synthetic observations are constructed directly from the emission-line model grid using a set of “true” physical parameters $\log(Z_\text{ISM})$, $\log(U)$, and $\log(n_\text{H})$. These are sampled continuously across the entire model grid, rather than from the discrete grid nodes themselves, thereby testing the accuracy of the interpolation scheme throughout the model parameter space. 
    To emulate observational data, fractional uncertainties are derived from the average SNR of the N=16 sample of LAEs. Gaussian noise is then applied by perturbing each emission-line flux with a random draw from a normal distribution with mean equal to the model-predicted value ($\mu = F_{\text{model}}$) and standard deviation set by the assigned uncertainty ($\sigma = \sigma_{\text{line}}$), producing mock observations. 
    The perturbed emission line fluxes are then fit using the same grid-based inference framework applied to the observational dataset. The perturbation-and-fitting process is then repeated $N=1000$ times to characterize the scatter in the recovered values due to noise realizations, thereby providing an estimate of the method's intrinsic statistical uncertainty. \\ \\ This recovery test only quantifies the statistical uncertainty associated with measurement noise and interpolation within the adopted model framework, not accounting for systematic uncertainties that may arise from assumptions in the photoionization models such as spherical symmetry, the implemented abundance prescriptions based on \citet{gutkin_2016} or stellar population assumptions, such as the constant SFH over the last 100 Myr. \\ \\
    \noindent Fig. \ref{fig:grid_mc_1000} shows that the metallicity is generally well recovered, with a median offset from the $1:1$ relation of $-0.007$. However, grid values appear to be favored, although during sampling, model predictions at arbitrary points in parameter space are obtained via multilinear interpolation. This clustering around the grid nodes likely results from the limited resolution of the emission-line model grid and the use of piecewise-linear interpolation. It is difficult to disentangle the effects of $\log(n_\text{H})$ and $\log(U)$ \citep{Kewley_2002, Strom_2017} given only optical emission lines, particularly at high ionization parameters as expected in these targets. Thus, both the ionization parameter $\log(U)$ and the density $\log(n_\text{H})$ are not well recovered, with median offsets from the $1:1$ relation of $0.021$ and $-0.090$, respectively. \\ \\ 
    \noindent Commonly used diagnostic lines, such as R23 and O32, are degenerate in $\log(n_\text{H})$ and $\log(U)$ as seen in Fig. \ref{fig:r23_o32_grid_degeneracy}, making it difficult to determine these properties uniquely. This phenomenon also occurs when fitting only optical emission lines (see Tab. \ref{tab:emission_lines}) to the grid, which can be seen when examining a cornerplot, shown in Fig. \ref{fig:cornerplot}, of an example fit characteristic of the fitting results of targets for which metallicity is determined. There appears to be a pronounced anti-correlated degeneracy between the ionization parameter $\log(U)$ and density $\log(n_\text{H})$, indicating that these parameters are not independently constrained by the observed emission lines. This indicates that additional UV emission lines would be required to properly constrain the ionization conditions of the targets, and the lack of such data is likely responsible for the large uncertainties in recovered $\log(n_\text{H})$ and $\log(U)$ values when testing the model grid. However, optical emission lines such as $[\text{OII}]$ and $[\text{NII}]$ can still be used to accurately determine metallicity, as they are only weakly dependent on the ionization parameter and density \citep{Kewley_2002}. For this reason, when testing the model grid, metallicity remains well constrained even when $\log(n_\text{H})$ and $\log(U)$ are not. Thus, metallicity determinations are likely to be indicative of the true gas-phase metallicity given high-quality observations of optical emission lines, whereas no precise statements can be made about the ionization parameter and gas density.
\section{ISM properties}\label{sec:ISM_prop}
    %\textit{no clear trends with Ly$\alpha$ EW, but large errors due to limited number of lines (see discussion). Also include stacks! Balmer decrement lower than theoretical value in some systems} 
    \begin{figure}
        \centering
        \includegraphics[width=\linewidth]{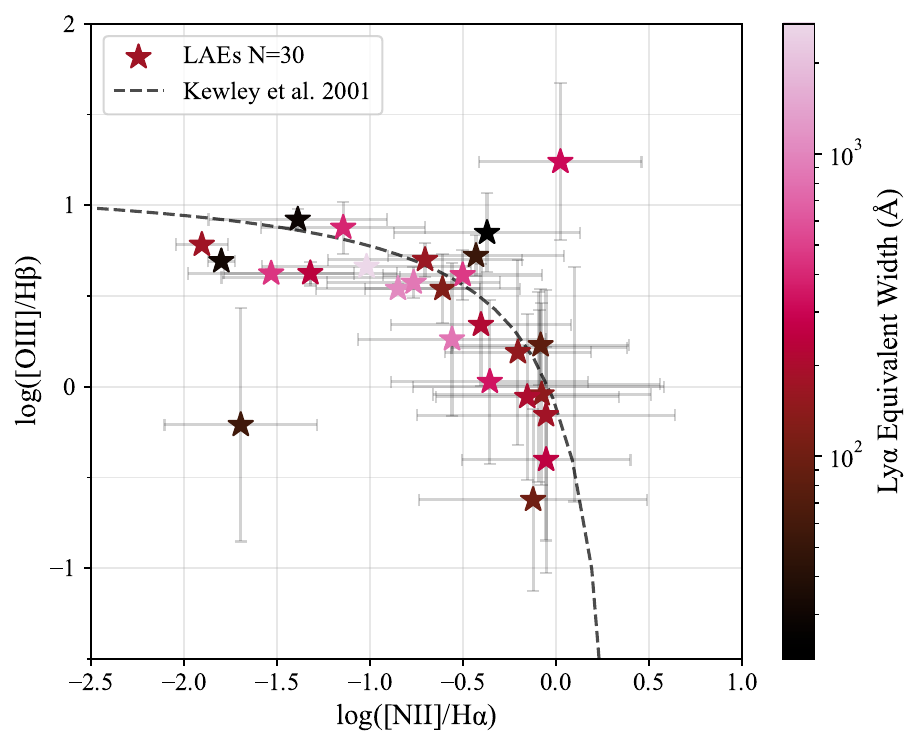}
        \caption{BPT diagram showing $\log([\mathrm{OIII}]\lambda5007/\mathrm{H}\beta)$ versus $\log([\mathrm{NII}]\lambda6584/\mathrm{H}\alpha)$ for the LAE sample with $\geq3$ emission lines detected in NIRSpec $(N=30)$. Points are color-coded by their Ly$\alpha$ EW, with error bars indicating measurement uncertainties. The dashed curve shows the theoretical maximum starburst line from \citet{kewley_2001}. Sources predominantly lie in the star-forming region, consistent with low metallicity and high ionization conditions.}
        \label{fig:bpt_by_ew}
    \end{figure} 
    \begin{figure}
        \centering
        \includegraphics[width=\linewidth]{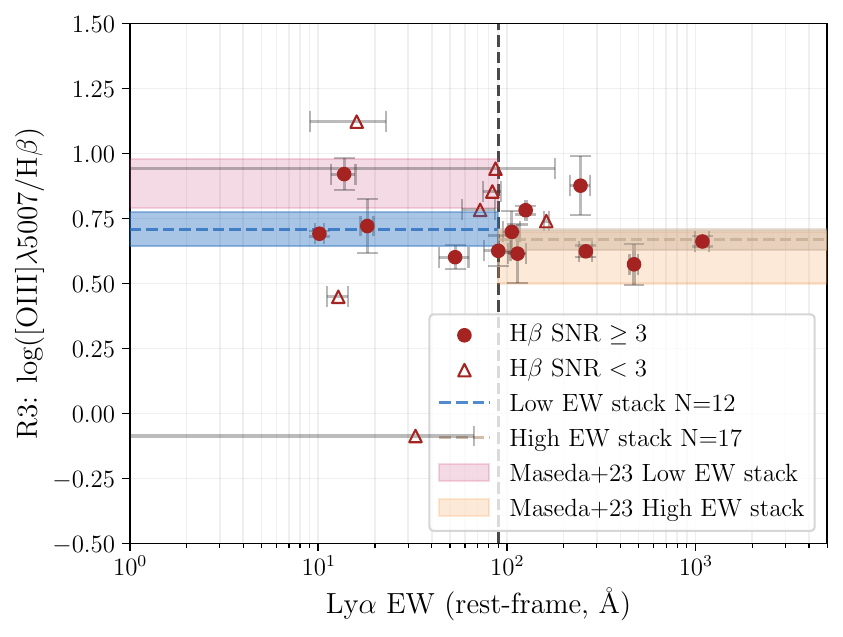}
        \caption{R3 ($\log([\text{OIII}]\lambda5007/\text{H}\beta)$) diagnostic as a function of Ly$\alpha$ EW. The sample of LAEs with secure line measurements (N=20) is divided into sources with a secure H$\beta$ detection (i.e., H$\beta$ SNR $\geq3$) and upper limits on H$\beta$ (H$\beta$ SNR $<3$). The mean value of R3 appears to decrease with EW when the sample is divided into low-EW (EW < 90 \AA) and high-EW (EW $\geq$ 90 \AA) subsets.}
        \label{fig:R3_vs_EW}
    \end{figure}
    \begin{figure}
        \centering
        \includegraphics[width=\linewidth]{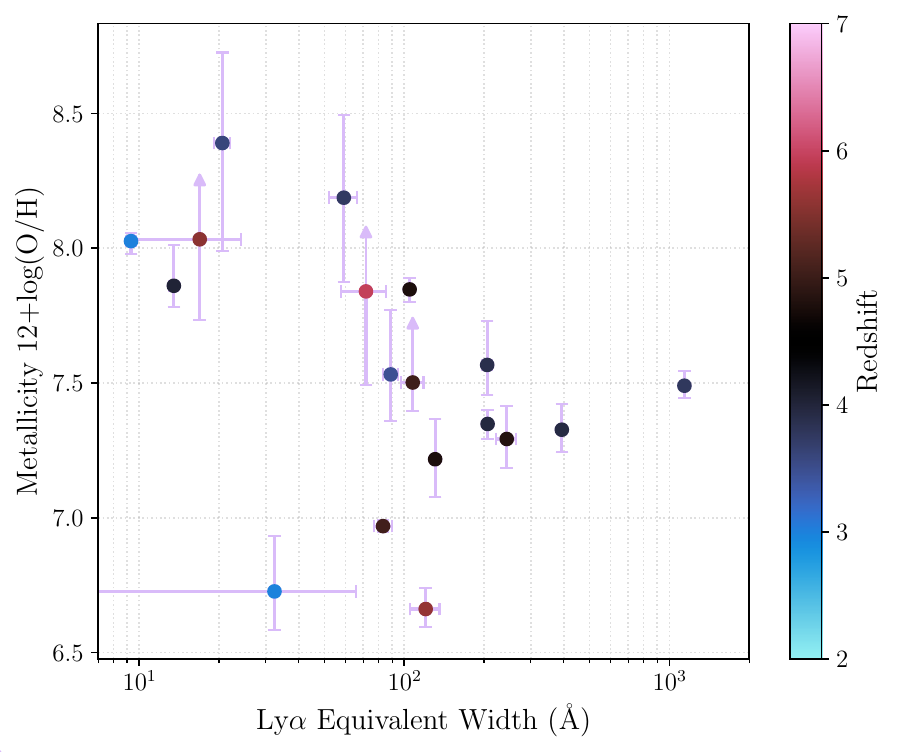}
        \caption{Gas-phase metallicity ($12+\log(\text{O/H})$) as a function of Ly$\alpha$ equivalent width (EW) for the LAE sample. Points are color-coded by redshift, with error bars reflecting uncertainties from the emission-line grid fitting. While a linear correlation is not statistically significant (Pearson $r=-0.12$, $p=0.51$), a moderate monotonic trend is suggested by the Spearman rank correlation ($\rho=-0.56$, $p=0.015$).}
        \label{fig:Z_vs_EW_by_redshift}
    \end{figure}
    %\begin{figure}
        %\centering
        %\includegraphics[width=\linewidth]{figures/o3hb_vs_z_by_nearest_u_and_n_to_bian.pdf}
        %\caption{R3 ($\log([\text{OIII}]/\text{H}\beta)$) as a function of metallicity for the BPASS emission-line grid. Model predictions are shown for different ionization parameters ($\log(U)=-3.5$ and $-2.5$) and span a range of gas densities (color-coded by $\log(n_\text{H})$). The curve shows the calibration from \citet{Bian_2018}, validated over $7.8\lesssim12+\log(\text{O/H})\lesssim8.4$. Significant deviations from the calibration are evident at low metallicities.}
        %\label{fig:R3_for_grid}
    %\end{figure}
    \begin{figure}
        \centering
        \includegraphics[width=\linewidth]{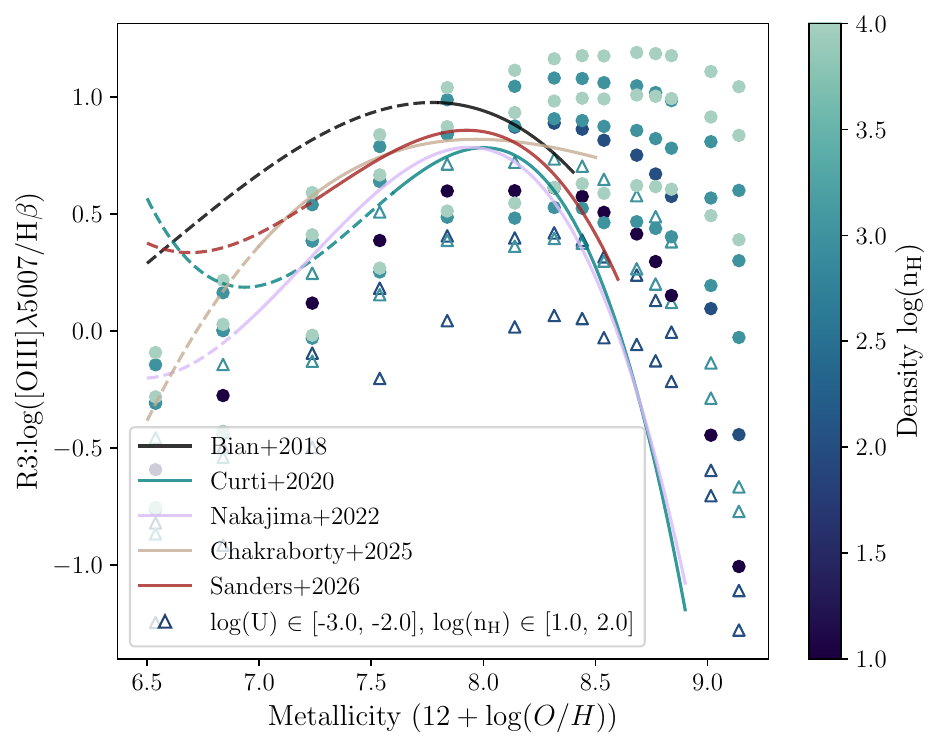}
        \caption{R3 ($\log([\text{OIII}]\lambda5007/\text{H}\beta)$) as a function of metallicity for the BPASS photoionization model grid. Model predictions are shown for ionization parameters $\log(U)=-3.0,-2.0,-1.0$ and hydrogen number densities $\log(n_\text{H})=1.0, 2.0, 3.0, 4.0$. Triangle markers indicate the grid parameters akin to the conditions present in \citet{Bian_2018}'s calibration. Curves show several empirical and JWST-era strong-line calibrations from \citet{Bian_2018}, \citet{curti_mzr_fmr_2020}, \citet{Nakajima_2022}, \citet{Chakraborty_2025}, and \citet{Sanders_2026}. While broad agreement is found at intermediate metallicities, substantial divergence is evident in the low-metallicity regime occupied by many LAEs, highlighting the calibration dependence of R3-based metallicity estimates under extreme ISM conditions.}
        \label{fig:comp_strong_line_calib}
    \end{figure}
    %\begin{figure}
        %\centering
        %\includegraphics[width=\linewidth]{figures/o3hb_vs_z_by_nh_with_other_r3.pdf}
        %\caption{R3 ($\log([\text{OIII}]\lambda5007/\text{H}\beta)$) as a function of metallicity for the BPASS photoionization model grid. Model predictions are shown for ionization parameters $\log(U)=-3.0,-2.0,-1.0$ and hydrogen number densities $\log(n_H)=1.0, 2.0, 3.0, 4.0$ Curves show several empirical and JWST-era strong-line calibrations from \citet{Bian_2018}, \citet{curti_mzr_fmr_2020}, \citet{Nakajima_2022}, \citet{Sanders_2026}, and \citet{Chakraborty_2025}. While broad agreement is found at intermediate metallicities, substantial divergence is evident in the low-metallicity regime occupied by many LAEs, highlighting the calibration dependence of R3-based metallicity estimates under extreme ISM conditions.}
        %\label{fig:comp_strong_line_calib}
    %\end{figure}
    Fig. \ref{fig:bpt_by_ew} shows that generally targets occupy regions of the BPT diagram consistent with highly star-forming systems, with many sources straddling the classical demarcation between star-forming galaxies and AGN as defined by \citet{kewley_2001} in the BPT. The elevated $[\text{OIII}]\lambda5007/\text{H}\beta$ ratios observed across the sample indicate high ionization conditions, consistent with hard ionizing radiation fields produced by young, low-metallicity stellar populations, as expected for a population of Ly$\alpha$ selected systems. 
    \\ \\ While such high ionization can, in principle, also be associated with AGN activity, there is no compelling evidence for the presence of AGN activity in the here presented targets. Galaxies are selected to be continuum-faint and exhibit only narrow emission lines, both of which are characteristic of star-forming systems and disfavor the presence of luminous, broad-line AGN \citep{Gawiser_2007, Feltre_2016}. This interpretation is further supported by recent JWST results, which show that high-redshift galaxies commonly exhibit elevated ionization parameters and line ratios overlapping with classical AGN regions even in the absence of clear AGN activity, driven instead by their low metallicities and extreme star-forming conditions \citep{Scholtz_2025b}. The systematically low $[\text{NII}]\lambda6584/\text{H}\alpha$ ratios further support a metal-poor nature for these galaxies, in line with expectations for Ly$\alpha$-selected systems, though they exhibit large uncertainties due to only faint detections of $[\text{NII}]$.
    \\ \\ Notably, no clear trends are observed between their position in the BPT and the Ly$\alpha$ equivalent width, suggesting that the high levels of star formation and ionization are not directly correlated with the observed Ly$\alpha$ EW, suggesting that the processes governing Ly$\alpha$ escape are largely decoupled from the nebular ionization conditions traced by these line ratios. \\ \\  Fig. \ref{fig:R3_vs_EW} shows the emission line diagnostic R3 ($\log([\text{OIII}]\lambda5007/\text{H}\beta)$) as a function of Ly$\alpha$ EW. Stacks from \citet{maseda_2023} show a significant step between the low EW sample ($<90$ \AA) and the high EW sample ($\geq90$ \AA); the step diminishes for the stacks produced in this work, as the low EW stack shows a lower line ratio. When testing for a linear correlation, even when taking into account the errors on the R3 measurements, we find no statistically significant trend with Pearson p-values $p=0.4665$ and Spearman p-values $p=0.3813$. We also test whether there is a significant difference between the low EW and high EW samples by performing an independent t-test between the two populations of LAE. The mean difference in R3 between the two samples is merely $0.0957$ dex, and the t-test shows no significance with a p-value of $p=0.3183$. \\ \\ The tendency for higher Ly$\alpha$ EW LAEs to have lower metallicity is also not preserved in this sample when metallicities are inferred via full emission-line grid fitting, taking into account all emission lines present in the spectra, as shown in Fig. \ref{fig:Z_vs_EW_by_redshift}. While previous studies have reported an anti-correlation between Ly$\alpha$ EW and metallicity, a linear correlation is not statistically significant (Pearson $r=-0.12$, $p=0.51$). However, a moderate monotonic trend is suggested by the Spearman rank correlation ($\rho=-0.56$, $p=0.015$). The large scatter at fixed EW,  with galaxies spanning over an order of magnitude in metallicity at $\text{EW}_{0,\text{ Ly}\alpha} \sim 100-300$ \AA, indicates that any dependence of metallicity on Ly$\alpha$ EW is weak and not well described by a simple linear relation. Moreover, any apparent weak trend is further diluted when considering the redshift distribution, suggesting that selection effects or evolving ISM conditions may be at play. The independent t-test also does not indicate a significant difference between the low- and high-EW population with regard to their gas-phase metallicity with a p-value of $p=0.1336$. Additionally, given the small sample size ($N=20$), we do not consider this evidence for a robust metallicity-EW relation. %As such, we are unable to confirm the relationship reported by \cite{maseda_2023} within this sample. 
    \\ \\ This apparent discrepancy with previous work may in part arise from the different techniques used to determine the Ly$\alpha$ equivalent widths of targets. Because targets were specifically selected to be continuum-faint in HST imaging, deep NIRCam data allow for tighter, or at least comparable, EW constraints than previous work. JADES in GOODS-S reaches depths of m$_\text{AB}\sim29.66\text{ mag}$ for point sources in F090W using NIRCam, while F160W imaging with WFC3 reaches m$_\text{AB}\sim27.21\text{ mag}$ for point sources \citep{Robertson_2026}, though depths in the eXtremely Deep Field (XDF) may be more comparable. Another possible explanation for this difference lies in the method of determining metallicity. The R3 diagnostic from \cite{Bian_2018}, used to infer metallicities in \citet{maseda_2023}, is calibrated on local analogs of $z\sim2$ star-forming galaxies and may not be directly applicable to the physical conditions present in high-redshift LAEs, as ISM conditions may evolve with redshift. In particular, these systems are believed to exhibit elevated ionization parameters and harder ionizing spectra, which can bias strong-line calibrations. More recent strong-line calibrations from JWST spectroscopy of high-redshift galaxies, such as \citet{Sanders_2026}, are better suited to these conditions and show greater consistency with the emission-line grid as seen in Fig. \ref{fig:comp_strong_line_calib}, which shows the behavior of the R3 diagnostic over a range of gas-phase metallicities. The relation between R3 and metallicity is clearly not one-to-one, but instead exhibits a strong dependence on ionization parameter $\log(U)$ and gas density $\log(n_\text{H})$ (parameters were chosen to mimic conditions derived for the targets in \cite{Bian_2018} and are shown as triangles). Furthermore, the calibration begins to break down at low metallicities ($Z \lesssim 0.1\text{ Z}_\odot$, i.e. $12+\log(\text{O/H}) \lesssim 7.7$), where the model grid predictions systematically deviate from the \cite{Bian_2018} relation. This is expected, as the calibration is extrapolated below its nominal validity range ($Z \gtrsim 0.13 \text{ Z}_\odot$ or $12+\log(\text{O/H}) \gtrsim 7.8$), leading to increasing uncertainty for the most metal-poor systems in the sample. To assess whether the discrepancies are unique to the \citet{Bian_2018} calibration, we compare the grid to several recent strong-line calibrations derived using JWST spectroscopy in Fig. \ref{fig:comp_strong_line_calib}. While broad agreement is found at intermediate metallicities, substantial divergence emerges in the low-metallicity regime relevant to many LAEs. The spread in model predictions at fixed metallicity reflects the sensitivity of R3 to both ionization parameter and gas density. Although the grid differs significantly from the \citet{Bian_2018} relation at low metallicity, it exhibits closer agreement with more recent high-redshift calibrations, particularly those of \citet{Nakajima_2022}, \citet{Chakraborty_2025} and \citet{Sanders_2026}, for models with higher gas densities. This suggests that part of the discrepancy may arise from differences in the physical conditions assumed by the various calibrations. Nevertheless, metallicity estimates based solely on R3 remain calibration-dependent and are further subject to systematic uncertainties associated with the adopted photoionization models, including assumptions regarding the ionizing spectrum, nebular structure, and chemical abundance patterns. \\ \\
    Beyond the limitations of the R3 diagnostic specifically, strong-line metallicity indicators more broadly have been shown to deviate systematically from direct electron temperature ($T_e$) based inferences, particularly in the low-metallicity, high-ionization regime occupied by many LAEs (see e.g. \citet{Nakajima_2022, Laseter_2024a}). The full emission line fitting approach adopted here, which simultaneously constrains multiple line fluxes across a broad wavelength range, is better positioned to break the degeneracies inherent to single-ratio diagnostics like R3. By fitting all available emission lines self-consistently, the derived physical parameters are more tightly constrained and less sensitive to the particular assumptions that limit any individual strong-line calibration.

    %\textit{Some of the recovered UV continuum slopes fall around $\beta\simeq-3.0$, which is an indicator of an extremely young and metal-poor stellar population \citep{Bouwens_2014, Chisholm_2022} and a large ionizing photon escape fraction \citep{Robertson_2010}.} 
    %\textit{what do we trust? can the stellar masses be trusted? what about the estimation of SFR from h$\alpha$?}
    
\section{Discussion}\label{sec:res_disc}
\subsection{Selecting metal-poor systems}
    %\begin{figure*}
        %\centering
        %\includegraphics[width=\linewidth]{figures/MZR_with_deltaMZR_vs_EW_gray.pdf}
        %\caption{Mass–metallicity relation for galaxies at $z \simeq 2-10$. Stellar mass ($\log M_\star/M_\odot$) is plotted against gas-phase metallicity (12 + log(O/H)). Gray circles and crosses show comparison samples from JADES \citep{Curti_2024} and CEERS \citep{nakajima_2023}, respectively. Star symbols denote the galaxies in this work, color-coded by Ly$\alpha$ equivalent width (N=16). Residuals relative to the MZR relation found by \citet{Curti_2024} decrease with increasing EW.  The lower panels show $\Delta$MZR as a function of EW with a linear fit yielding $R=-0.55$ and a one-sided $p_{1s}=0.014$ for the hypothesis of a negative trend.}
        %\label{fig:mzr}
    %\end{figure*}
    Our sample of LAEs spans a broad range in metallicity at fixed Ly$\alpha$ EW, with substantial scatter across the full sample. This suggests that Ly$\alpha$ EW alone may not be a reliable predictor of gas-phase metallicity in this population. However, the moderate monotonic trend suggested by the Spearman rank correlation between Ly$\alpha$ EW and metallicity indicates that there may still be a second-order trend linking Ly$\alpha$ strength to nebular conditions, even if it does not translate into a simple linear relation with metallicity.\\ \\ One possible interpretation is that Ly$\alpha$ escape and gas-phase enrichment are only indirectly coupled. Ly$\alpha$ emission is strongly sensitive to radiative transfer effects, geometry, and neutral hydrogen distribution, while metallicity traces the chemical enrichment of galaxies. As a result, variations in ISM structure, covering fraction, or kinematics may dominate over metallicity in regulating the observed Ly$\alpha$ EW, particularly in low-mass, highly ionized systems where feedback-driven outflows can significantly modify the neutral gas distribution. \\ \\ A further complication arises from the distinction between stellar and gas-phase metallicity, a distinction that this work implicitly sets aside by treating them as equivalent. However, while gas-phase metallicity, typically traced by the oxygen abundance O/H, reflects the current chemical state of the nebular gas, stellar metallicity, traced by [Fe/H], can directly affect the ionizing continuum. Massive stars with lower [Fe/H] produce harder ionizing spectra, which boosts the intrinsic Ly$\alpha$ production rate and, in turn, the intrinsic Ly$\alpha$ EW \citep{Topping_2020}. A correlation between Ly$\alpha$ EW and stellar metallicity may therefore exist even without a tight relation to gas-phase metallicity. This decoupling can arise naturally from variations in the $\alpha$ element abundances: in young, rapidly star-forming systems, $\alpha$ element enrichment from core-collapse supernovae can proceed on much shorter timescales than iron enrichment, leading to a situation where low [Fe/H] does not necessarily correspond to low O/H. Consequently, the scatter observed in Fig. \ref{fig:Z_vs_EW_by_redshift} between Ly$\alpha$ EW and gas-phase metallicity may in part reflect genuine variation in the $\alpha$ abundances across the sample. Disentangling these contributions is currently beyond the reach of the available data, as targets are sufficiently continuum-faint that the stellar continuum remains undetected, leaving only nebular emission lines accessible.\\ \\ The distribution of galaxies in the mass–metallicity plane broadly follows the established relation at high redshift, showing general consistency with the trends reported by \cite{Curti_2024} and \ \cite {nakajima_2023}. However, a systematic offset is evident for galaxies with high Ly$\alpha$ EWs, which tend to fall below the mean relation. The residuals relative to the best-fit MZR relation of \citet{Curti_2024} exhibit evidence for an anti-correlation with Ly$\alpha$ equivalent width. A Pearson test yields $R = -0.55$ ($p_{1s} = 0.014$), while a Spearman rank test gives $\rho = -0.594$ ($p = 0.0152$). The trend remains negative under leave-one-out resampling, indicating that it is not driven by any individual galaxy in the sample. While the sample size remains modest (N = 16), these results suggest that galaxies with stronger Ly$\alpha$ emission appear to preferentially occupy lower metallicities than expected for their stellar mass, consistent with scenarios in which they represent less chemically evolved systems or are experiencing recent inflows of metal-poor gas. Although no direct correlation is observed between metallicity and Ly$\alpha$ EW (Fig. \ref{fig:Z_vs_EW_by_redshift}), a significant anti-correlation emerges when considering residuals relative to the mass–metallicity relation shown in Fig. \ref{fig:mzr_comp_SPHINX}. This suggests that stellar mass is a major source of scatter in the metallicity distribution and may obscure weaker secondary dependencies on Ly$\alpha$ EW. Thus, combining Ly$\alpha$ EW with stellar mass may provide an effective alternative selection criterion for identifying the most metal-poor systems, potential analogs of the first galaxies that contributed to cosmic reionization.
\subsection{Comparison to Simulations}
    %\begin{figure}
        %\centering
        %\includegraphics[width=\linewidth]{figures/lya_ew_histogram_obs_ew.pdf}
        %\caption{Normalized histogram of Ly$\alpha$ equivalent widths for the galaxies included in this study (N=16) compared to a sample of \texttt{SPHINX$^{20}$} targets in the redshift range $4\leq z\leq6$ from \citet{Katz_2023a} (N=960). The observed Ly$\alpha$ equivalent width distribution closely matches the central portion of the \texttt{SPHINX$^{20}$} distribution, enabling a meaningful comparison between the two.}
        %\label{fig:ew_comp_SPHINX}
    %\end{figure}
    \begin{figure}
        \centering
        \includegraphics[width=\linewidth]{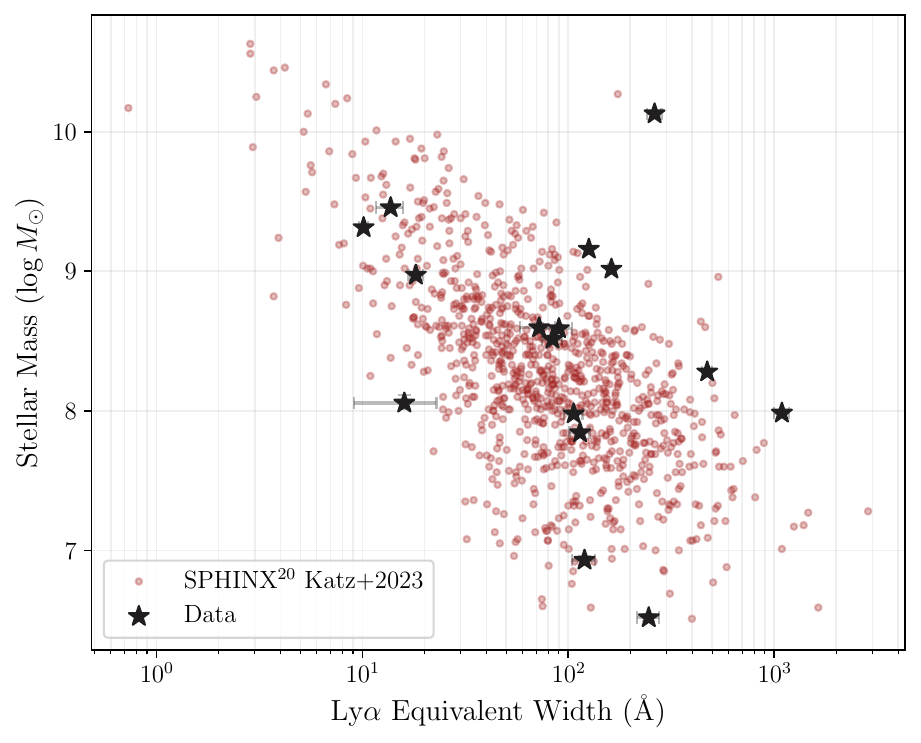}
        \caption{Stellar masses and equivalent widths of the LAE sample ($N=16$) compared to a sample of \texttt{Sphinx$^{20}$} targets in the redshift range $4\leq z\leq6$ from \citet{Katz_2023a} (N=960). The observed distribution of Ly$\alpha$ equivalent widths and stellar masses matches the distribution of galaxies selected from \texttt{SPHINX$^{20}$}, enabling a meaningful comparison between the two.}
        \label{fig:ew_comp_SPHINX}
    \end{figure}
    \begin{figure}
        \centering
        \includegraphics[width=\linewidth]{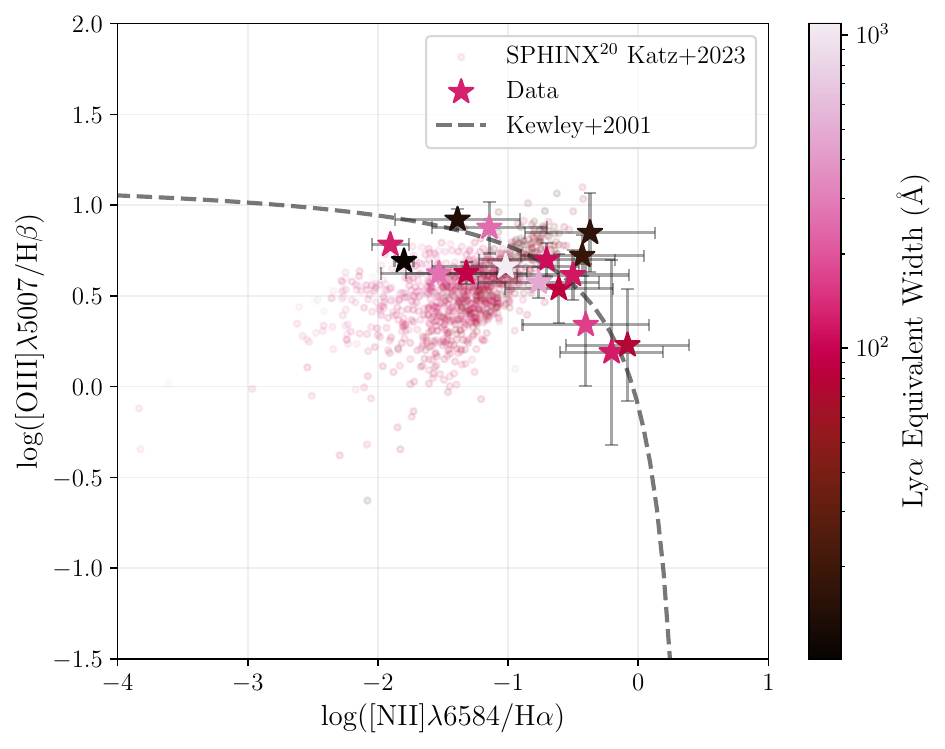}
        \caption{Comparison of the nebular emission-line ratios of the observational sample (star symbols) and \texttt{SPHINX$^{20}$} galaxies (circles) in the BPT diagram. Points are colored by Ly$\alpha$ EW, and the \citet{kewley_2001} theoretical star-forming boundary is shown as the dashed curve. Although the simulations overlap substantially with the observed distribution, they occupy a narrower locus. This suggests that, while \texttt{SPHINX$^{20}$} captures the general ionization properties of high-redshift LAEs, it does not fully reproduce the most extreme nebular conditions inferred for the observed sample.}
        \label{fig:bpt_comp_SPHINX}
    \end{figure}
    \begin{figure*}
        \centering
        \includegraphics[width=\textwidth]{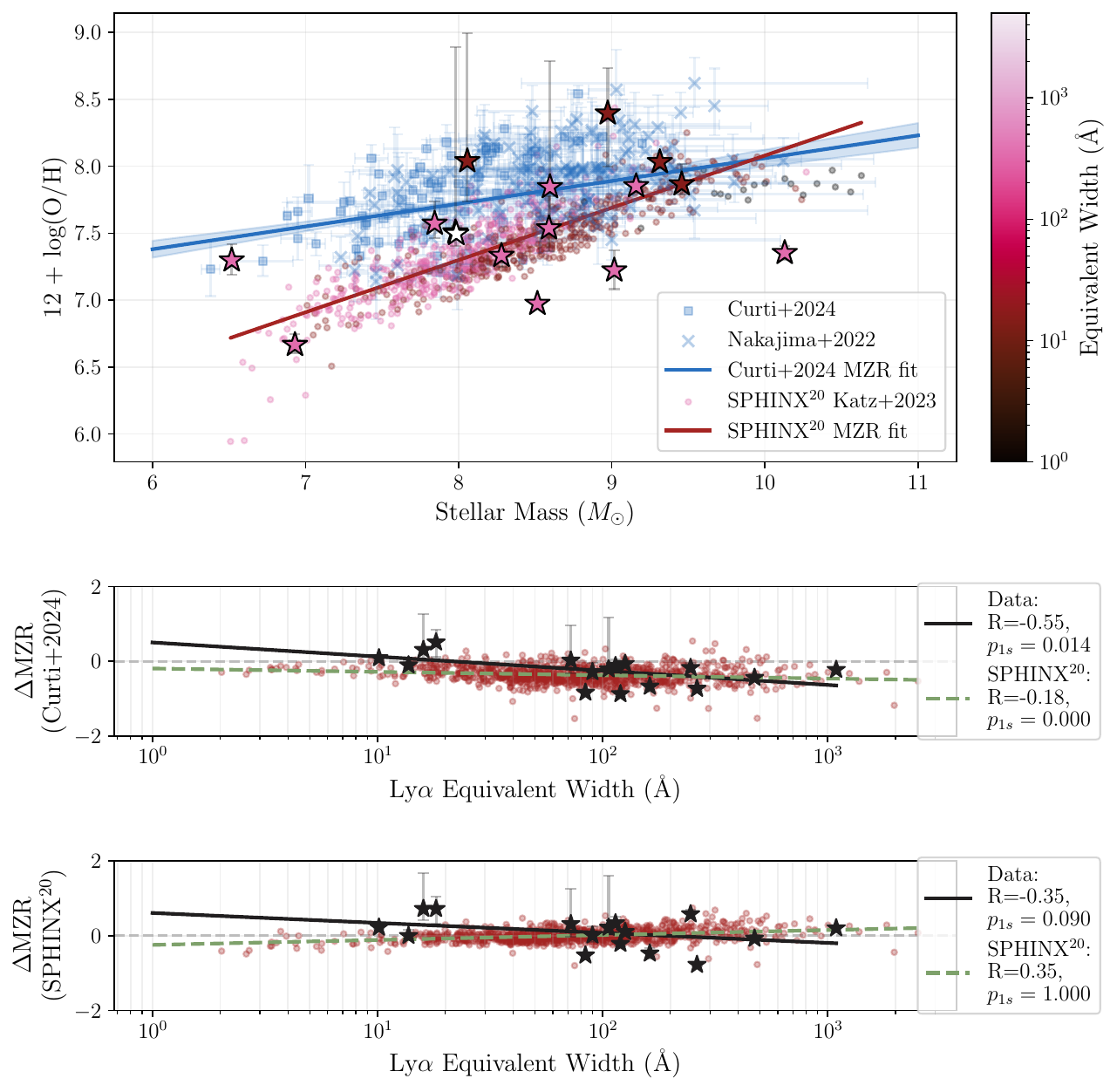}
        \caption{Comparison between the observed sample and the \texttt{SPHINX$^{20}$} simulations in stellar mass, metallicity, and Ly$\alpha$ equivalent width. Top: Stellar mass-metallicity relation for the observations (black stars) and \texttt{SPHINX$^{20}$} galaxies (red). The JADES sample and empirical mass-metallicity relation from \citet{Curti_2024} (blue) and the best-fitting relation to the \texttt{SPHINX$^{20}$} sample (red line) are shown for reference. While both samples span a similar stellar mass range, the simulated galaxies follow a steeper mass-metallicity relation. Middle: Offset from the \citet{Curti_2024} mass-metallicity relation, $\Delta$MZR, as a function of Ly$\alpha$ equivalent width. Both the observations and \texttt{SPHINX$^{20}$} exhibit negative trends with increasing EW. Bottom: $\Delta$MZR measured relative to the best-fitting \texttt{SPHINX$^{20}$} mass-metallicity relation. In this case, the simulations exhibit a positive correlation with Ly$\alpha$ equivalent width, whereas the observational sample remains anti-correlated, illustrating that inferred metallicity residual trends depend sensitively on the adopted reference mass-metallicity relation.}
        \label{fig:mzr_comp_SPHINX}
    \end{figure*}
    To place our results in the context of current theoretical models, we compare our sample to forward-modeled galaxies from the \verb|SPHINX|$^{20}$ simulations \citep{Katz_2023a}. \verb|SPHINX| \citep{Rosdahl_2018}, and its successor \verb|SPHINX|$^{20}$ \citep{Rosdahl_2022} are cosmological radiation-hydrodynamics simulations, simultaneously modeling the large-scale process of cosmic reionization and the detailed physics of a multiphase ISM, providing a statistical sample of galaxies akin to those currently being observed by JWST \citep{Katz_2023a}. The simulations reproduce several observed statistical properties of galaxies during the epoch of reionization, including the UV and Ly$\alpha$ luminosity functions over the mass and luminosity range probed by \verb|SPHINX|$^{20}$ \citep{Garel_2021}, and provide mock galaxy populations designed for direct comparison with JWST observations making them a useful benchmark for interpreting the physical origin of the trends identified in this work \citep{Katz_2023a}. \\ \\ To ensure a fair comparison, we consider only simulated galaxies occupying a similar redshift range as our observations, $2\leq z\leq6$, limiting the sample of simulated galaxies to redshifts $z\in[4.64, 5.0, 6.0]$. Figure \ref{fig:ew_comp_SPHINX} shows that the \verb|SPHINX|$^{20}$ subsample of galaxies is a good match in stellar mass, with the Ly$\alpha$ equivalent width distribution also closely matching that of our observational sample, indicating that both datasets probe a comparable population of Ly$\alpha$ emitters. \\ \\ Despite these similarities, several systematic differences are evident. Figure \ref{fig:bpt_comp_SPHINX} shows the BPT diagram, in which the \verb|SPHINX|$^{20}$ subsample occupies a narrower locus and does not extend into the region populated by a number of the observed LAEs, particularly for higher values of $\log([\mathrm{NII}]/\mathrm{H}\alpha$). This suggests that the simulations do not fully reproduce the most extreme nebular conditions inferred for some observed LAEs, potentially reflecting differences in ionization parameter, nitrogen abundance, stellar ionizing spectra, or unresolved ISM structure.\\ \\ Differences are also apparent in the mass-metallicity relation. While both samples exhibit increasing metallicity with stellar mass, the \verb|SPHINX|$^{20}$ relation is considerably steeper than that inferred from the observational calibration using data obtained during the JWST Advanced Deep Extragalactic Survey (JADES) \citep{Eisenstein_2026, Curti_2024}. Consequently, the simulated galaxies are systematically more metal-poor than galaxies observed with JWST across a redshift range of $3\leq z \leq 10$, especially at low stellar masses relative to the empirical relation. The observational sample spans both sample distributions, highlighting the sensitivity of any metallicity residual analysis to the adopted reference MZR. \\ \\ This dependence is reflected in the trends between Ly$\alpha$ equivalent width and metallicity offset shown in the lower panels of Fig. \ref{fig:mzr_comp_SPHINX}. When offsets are measured relative to the \citet{Curti_2024} relation, both the observations and the \verb|SPHINX|$^{20}$ galaxies sample exhibit negative trends with increasing Ly$\alpha$ EW. However, when offsets are instead defined relative to the best-fitting \verb|SPHINX|$^{20}$ mass-metallicity relation, the simulations show a positive correlation, whereas the observational sample remains anti-correlated, though with less statistical significance. Thus, the apparent relationship between Ly$\alpha$ equivalent width and metallicity residual may not be universal, but depends strongly on the underlying reference mass-metallicity relation. \\ \\ Several factors may contribute to the observed discrepancies. On the observational side, the metallicity estimates for several galaxies carry substantial uncertainties. Current high-redshift JWST spectroscopic samples are subject to selection effects that preferentially target galaxies with strong continuum emission or strong emission lines. Consequently, JWST high-redshift samples such as JADES \citep{Curti_2024} and CEERS \citep{nakajima_2023} may not be fully representative of the entire galaxy population at high redshift. On the theoretical side, although \verb|SPHINX|$^{20}$ incorporates radiation hydrodynamics and Ly$\alpha$ radiative transfer, uncertainties in the modelling of stellar populations, chemical enrichment, stellar feedback, and the unresolved multiphase ISM can influence both the predicted nebular line ratios and Ly$\alpha$ escape. 
    %Furthermore, the largest Ly$\alpha$ equivalent widths may be enhanced by orientation-dependent radiative transfer effects, whereby dust preferentially suppresses the UV continuum while resonantly scattered Ly$\alpha$ photons are redirected into the observer's line of sight. 
    Finally, the limited number of LAEs with both robust metallicity measurements and stellar mass estimates currently restricts the statistical power of the comparison. Future JWST observations spanning a broader range of galaxy properties will be essential for establishing whether these differences reflect limitations of the simulations, observational selection effects, or genuine differences between the simulated and observed high-redshift galaxy populations. 
    %\\ \\ Overall, the comparison demonstrates that \verb|SPHINX|$^{20}$successfully reproduces the global properties of the galaxies targeted in this work, including their stellar masses and Ly$\alpha$ equivalent width distribution, but does not fully capture the nebular emission-line properties or the observed relationship between Ly$\alpha$ emission and metallicity. These differences highlight the need for further study of whether Ly$\alpha$ EW can be a useful proxy to identify the most metal-poor systems. The observational trends may also provide useful constraints for future generations of radiation-hydrodynamic simulations to match observations of LAEs.
\subsection{Limitations of Current Analysis}
    Forward-model testing of the emission-line grid-fitting highlights an intrinsic limitation of rest-frame optical emission-line diagnostics: metallicity is comparatively well constrained, but ionization parameter and gas density remain highly degenerate when only optical lines are available. This degeneracy, already noted in earlier studies (e.g., \cite{Kewley_2002, Strom_2017}), is exacerbated in the high-ionization regime typical of these systems. As shown by our recovery tests, even high-quality observations yield significant scatter in the recovery of $\log(U)$ and $\log(n_\mathrm{H})$, reinforcing the need for additional diagnostics beyond the optical regime. In particular, rest-frame ultraviolet emission lines such as $\text{HeII}\lambda1640$, $\text{CIV}\lambda1549$, and $\text{OIII]}\lambda\lambda1661,1666$ provide critical leverage on the hardness of the ionizing spectrum and ionization parameter, and are essential for breaking the degeneracies present in optical-only analyses and could be utilized to further refine our understanding of the nebular conditions of LAEs. \\ \\ Additionally, obtaining deeper rest-frame optical observations centered on the likely host galaxies, rather than the positions of the $\text{Ly}\alpha$ emission, could further improve metallicity constraints for these unique targets. By targeting the galaxy centers instead of the outskirts, these observations would increase the number of reliable measurements and likely provide better constraints on metal lines such as $[\text{NII}]\lambda6548, 6584$ and $[\text{SII}]\lambda6717,6731$.\\ \\ In the future, the effects of varying the IMF and upper mass cut-off on emission line ratios should be investigated, especially given early JWST results showing an abundance of bright, massive galaxies at $z>10$, which may be explained by a top-heavy IMF \citep{Hutter_2024}. \\ \\ \verb|BPASS| models v2.3 \citep{Byrne_2022} additionally vary the amount of enhancement in $\alpha$ elements. An expansion of the grid of emission-line diagnostics to include the effect of varying $[\alpha/\text{Fe}]$ is planned. This is of particular interest as young stellar populations are predicted to show significant enhancement in $\alpha$ elements, as the timescale for $\alpha$ enrichment is shorter, originating from young, massive stars ending their life in core-collapse supernovae, while iron elements are introduced through supernovae of type II, occurring on longer timescales. 
    %\\ \\ In summary, high-EW LAEs reproduce several characteristics expected of reionization-era galaxies, including low stellar masses, low metallicities, and strong nebular emission. However, Ly$\alpha$ EW alone does not uniquely select for the most metal-poor systems, as our results do not support a simple monotonic relation between Ly$\alpha$ EW and metallicity. The apparent trends in stacked diagnostics likely reflect underlying population averages rather than a direct causal relationship at the level of individual galaxies. The results presented here are consistent with a more complex scenario in which selecting for the most metal-poor systems can be achieved when considering both Ly$\alpha$ EW and stellar mass. We therefore conclude that high-EW LAEs are useful, though imperfect, analogs of the galaxies believed to dominate reionization. 
\section{Conclusions}\label{sec:conclusions}
    In this work, we have investigated the ionizing and chemical properties of a sample of 88 Ly$\alpha$ emitters (LAEs) at intermediate redshifts ($2 \lesssim z \lesssim 7$). Combining JWST/NIRSpec spectroscopy with ancillary MUSE and NIRCam observations, we derived Ly$\alpha$ equivalent widths, stellar masses, and gas-phase metallicities using a grid of \verb|BPASS|+\verb|cloudy| photoionization models.\\ \\
    The primary results are as follows:
    \begin{enumerate}
        \item The LAEs in this sample occupy regions of optical emission-line diagnostic diagrams consistent with highly ionized, predominantly star-forming systems. Their elevated $[\text{OIII}]\lambda5007/\text{H}\beta$ ratios and generally low $[\text{NII}]\lambda6584/\text{H}\alpha$ ratios support the picture that these galaxies host young, metal-poor stellar populations. \\
        \item Forward-model tests demonstrate that gas-phase metallicities can be recovered reliably from the available optical emission lines, whereas the ionization parameter and hydrogen density remain strongly degenerate. Consequently, metallicity estimates are comparatively robust, while constraints on the detailed ionization conditions require additional rest-frame ultraviolet diagnostics.\\
        \item We find no compelling evidence for a strong linear relationship between Ly$\alpha$ equivalent width and gas-phase metallicity. The substantial scatter observed at fixed EW suggests that Ly$\alpha$ emission is influenced by additional factors beyond chemical enrichment alone. \\
        \item When stellar mass is taken into account, a different picture emerges. Galaxies with stronger Ly$\alpha$ emission tend to lie below the average mass--metallicity relation derived from observational data at comparable stellar masses, suggesting that high-EW LAEs are preferentially less chemically evolved than typical galaxies of the same mass. Although the current sample remains modest, this result indicates that Ly$\alpha$ EW in conjunction with stellar mass may be a good indicator for extremely metal-poor systems. \\
        \item Comparison with the \verb|SPHINX|$^{20}$ simulations shows that the observed LAEs occupy a similar stellar mass and Ly$\alpha$ equivalent width regime, but differ in their nebular emission-line properties and mass-metallicity relation. Given the modest sample size, larger spectroscopic samples spanning a wider range of galaxy properties will be essential to establish whether these discrepancies reflect the diversity of the high-redshift LAE population or are driven by selection effects
    \end{enumerate}
    Taken together, these results support the view that high-EW LAEs reproduce several key characteristics expected of the low-mass galaxies thought to dominate cosmic reionization, such as low metallicities and intense nebular emission. However, Ly$\alpha$ EW alone does not uniquely select the most metal-poor galaxies. Instead, the strongest Ly$\alpha$ emitters appear to be systems that are chemically under-enriched relative to their stellar mass, suggesting that low-mass, high-EW LAEs may indeed provide good analogs for the earliest galaxies. 
    %\\ \\ Future observations incorporating rest-frame ultraviolet emission lines and deeper spectroscopy targeted on the host galaxies themselves will be essential for constraining the ionization conditions of these systems and testing whether the trends identified here persist in larger samples. Nevertheless, the results presented in this work indicate that high-EW LAEs remain valuable, though imperfect, laboratories for studying the physical conditions of galaxies during the epoch of reionization.

\section*{Acknowledgements}
This work is based [in part] on observations made with the NASA/ESA/CSA James Webb Space Telescope. The data were obtained from the Mikulski Archive for Space Telescopes at the Space Telescope Science Institute, which is operated by the Association of Universities for Research in Astronomy, Inc., under NASA contract NAS5-03127 for JWST. These observations are associated with program \#1671. \\ \\ This work made use of v2.2.1 of the Binary Population and Spectral Synthesis (BPASS) models as described in \citet{eldridge_2017} and \citet{stanway_eldridge_2018}.
%%%%%%%%%%%%%%%%%%%%%%%%%%%%%%%%%%%%%%%%%%%%%%%%%%
\section*{Data Availability}
The data underlying this article are based on observations made with the NASA/ESA/CSA James Webb Space Telescope (JWST Program ID 1671, PI: Maseda), obtained from the Mikulski Archive for Space Telescopes (MAST) at the Space Telescope Science Institute and can be accessed via  \url{https://doi.org/10.17909/x1xc-k671}. \\ \\ This research also makes use of MUSE observations from the MUSE Ultra Deep Field (\citet{Bacon_2023}; \href{https://doi.org/10.18727/archive/85}{doi:10.18727/archive/85}) and the MUSE-Wide survey (\citet{Herenz_2017}; \citet{Urrutia_2019}; \href{https://doi.org/10.17876/musewide/dr.1/20}{doi:10.17876/musewide/dr.1/20} and \href{https://doi.org/10.17876/musewide/dr.1/21}{doi:10.17876/musewide/dr.1/21}), all publicly available from the ESO Science Archive Facility (\url{https://archive.eso.org}).



\bibliographystyle{mnras}
\bibliography{references} % if your bibtex file is called example.bib

%\begin{thebibliography}{99}
%\end{thebibliography}

%%%%%%%%%%%%%%%%%%%%%%%%%%%%%%%%%%%%%%%%%%%%%%%%%%

%%%%%%%%%%%%%%%%% APPENDICES %%%%%%%%%%%%%%%%%%%%%

\appendix

\section{LAEs with no clear host}

\begin{figure}
    \centering
    \includegraphics[width=\linewidth]{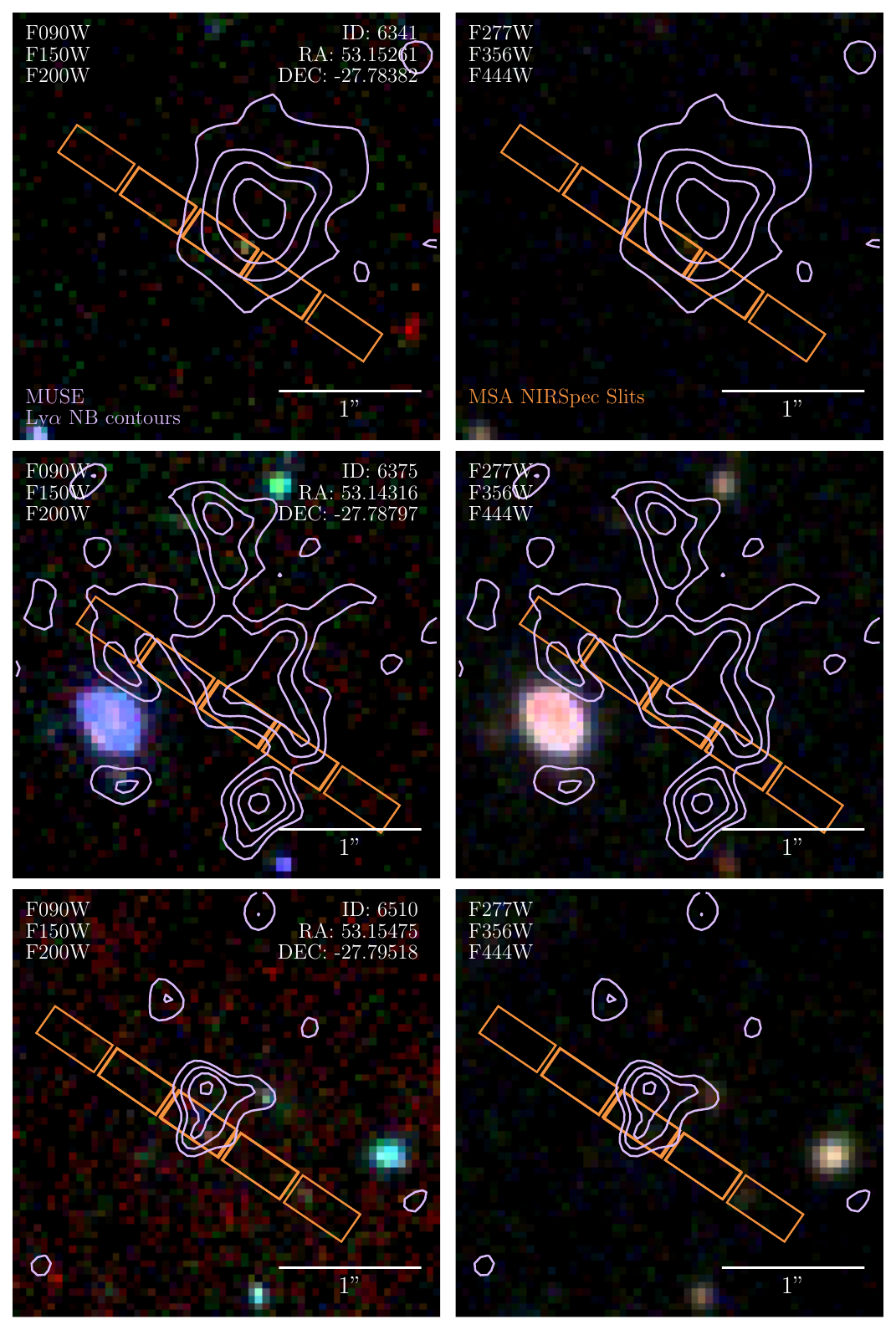}
    \caption{Examples of LAEs in the sample with no clear host associations: the top panels depict a blank field with no possible host visible; the middle panels show an example of several hosts within a 1 arcsecond radius whose photometric redshifts are consistent with the Ly$\alpha$ emission; the lower panels depict a faint host for which no photometric redshift is available from the DJA catalogs and thus cannot be clearly associated}
    \label{fig:non-detections}
\end{figure}

As detailed in Section \ref{sec:slit_positioning}, we only match hosts to Ly$\alpha$ emission if they are located within a 1 arcsecond radius and have a comparable photometric redshift with $\Delta z\leq0.1$. This leads to $N=38$ LAEs for which no clear association can be made, comprised of true non-detections as seen in the top panel of Fig. \ref{fig:non-detections}, LAEs with multiple possible hosts fulfilling the above criteria such as in the middle panel of Fig. \ref{fig:non-detections}, and faint possible hosts for which no redshift information is available in the photometric catalogs, shown in the bottom panels of Fig. \ref{fig:non-detections}. Out of these non-detections, only $N=4$ targets have no detection in NIRCam but show line emission in the MSA NIRSpec spectra. 

\section{Lyman alpha offsets}

\begin{figure}
    \centering
    \includegraphics[width=\linewidth]{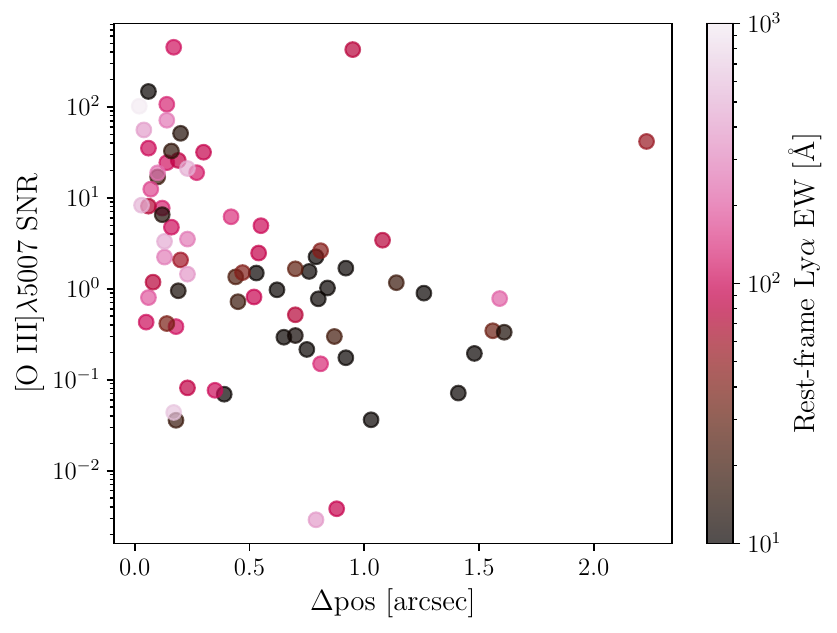}
    \caption{Offset between the Ly$\alpha$ emission position recorded by MUSE and the identified host in NIRCam imaging versus the SNR in $[\text{OIII}]\lambda5007$. As slits were positioned on the Ly$\alpha$ emission, smaller offsets tend to result in higher SNR in detected emission lines as more of the gas from the host galaxy is likely to have been captured during slit spectroscopy. There is no clear trend with Ly$\alpha$ EW, indicating that the here produced results are not biased by the survey design.}
    \label{fig:offsets_lya}
\end{figure}

We report the SNR of the $[\text{OIII}]\lambda5007$ emission line as a function of the projected offset between the Ly$\alpha$ emission measured from the MUSE observations and the associated host galaxy identified in the NIRCam imaging. Since the NIRSpec microshutters were centred on the Ly$\alpha$ coordinates derived from the MUSE data, the spatial offset between the Ly$\alpha$ emission and the stellar continuum directly affects the fraction of the host galaxy encompassed by the slit. Galaxies with smaller offsets are therefore expected to have a larger fraction of their nebular emission captured during the spectroscopic observations, resulting in higher SNR measurements of emission lines such as $[\text{OIII}]\lambda5007$. Conversely, larger offsets can lead to partial slit losses, reducing the observed line fluxes and their corresponding SNR. While slit losses are mostly negligible when evaluating line ratios and relative line strengths, no SFRs derived from H$\alpha$ are presented here as the slit losses can not be easily estimated thereafter. \\ \\
To assess whether this observational effect could bias our analysis, we also compare the Ly$\alpha$--NIRCam offset with the Ly$\alpha$ equivalent width. We find no evidence for a systematic relationship between offset and Ly$\alpha$ EW, indicating that galaxies with stronger Ly$\alpha$ emission are not preferentially associated with smaller positional offsets. This suggests that the trends presented throughout this work are not driven by the slit-placement strategy adopted for the NIRSpec observations, but instead reflect the intrinsic properties of the galaxy sample.

\section{Differences in EW measurements}

\begin{figure}
    \centering
    \includegraphics[width=\linewidth]{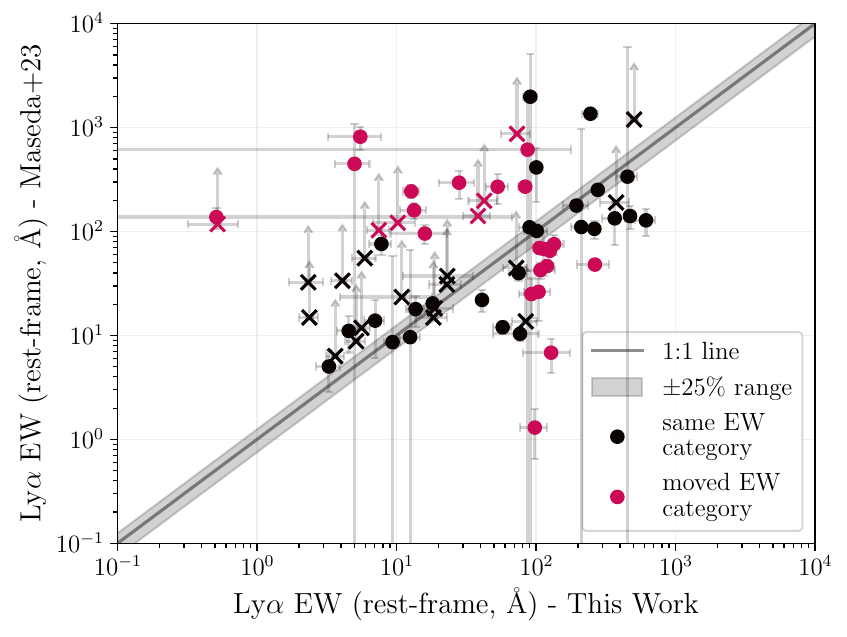}
    \caption{Ly$\alpha$ equivalent widths compared to the values utilized by \citet{maseda_2023}. LAEs are color-coded whether they change category (going from low EW ($<90$ \AA) to high EW ($\geq90$ \AA) and vice versa) or stay the same. Targets with only limits on the equivalent width from \citet{maseda_2023} are marked as crosses. A significant number of targets show large differences in the determined EWs, influencing the above-presented analysis and accounting for possible differences.}
    \label{fig:ew_comp_maseda}
\end{figure}

\begin{figure}
    \centering
    \includegraphics[width=\linewidth]{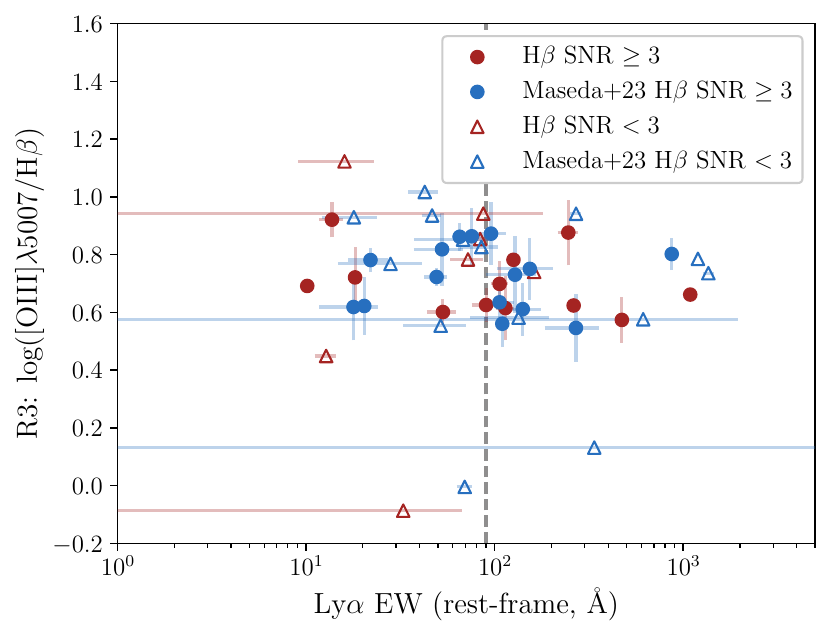}
    \caption{Fig. \ref{fig:R3_vs_EW} comparing this work's results to \citet{maseda_2023}. Though individual targets have been determined to exhibit alternate Ly$\alpha$ EWs and R3 values, the general distribution in the space is comparable. Using not the stacked analysis but individual targets of the sample with regard to identifying a linear trend with EW or a step between the population of low EW and high EW targets results in no statistically significant trends.}
    \label{fig:r3_vs_ew_comp_maseda}
\end{figure}

To assess the impact of our revised Ly$\alpha$ equivalent width (EW) measurements on the presented analysis, we compare our derived EWs with those reported by \citet{maseda_2023} in Fig.~\ref{fig:ew_comp_maseda}. While many galaxies exhibit good agreement, a substantial fraction show significant differences in their measured EWs. These revisions result in several objects changing between the low-EW ($<90$ \AA) and high-EW ($\geq90$ \AA) categories used throughout this work, demonstrating that the adopted EW measurements can influence sample classification and, consequently, the inferred population properties. \\ \\
To determine whether these revised measurements alter the overall conclusions, Fig.~\ref{fig:r3_vs_ew_comp_maseda} compares the relationship between Ly$\alpha$ EW and the R3 emission-line ratio using the measurements from this work and those of \citet{maseda_2023}. Although individual galaxies occupy different positions owing to the updated Ly$\alpha$ EWs and R3 measurements, the overall distribution of the sample remains similar between the two studies. Furthermore, considering the individual galaxies rather than the stacked spectra reveals neither a statistically significant linear correlation between R3 and Ly$\alpha$ EW nor evidence for a distinct step between the low- and high-EW populations. These results indicate that, despite revisions to individual measurements, the qualitative conclusions regarding the absence of a significant trend are robust.

\section{LAE Properties}

The following table summarizes the main properties of the 88 LAEs determined in this analysis, including Ly$\alpha$ EW, gas-phase metallicity, and stellar mass.

\onecolumn
\renewcommand{\TblrNewPage}{\clearpage}

    \begin{longtblr}[caption={Summary of LAE properties}, label={tab:lae_summary}]{colspec={X[c]|X[c]|X[c]|X[c]|X[c]|X[c]|X[c]}, rowhead=2, hline{2}={solid}, hline{3}={1}{-}{1pt}, hline{3}={2}{-}{1pt}}

\textbf{ID} & \textbf{RA} & \textbf{DEC} & \textbf{Redshift} & \textbf{Ly$\alpha$ EW} & \textbf{12 + log(O/H)} & \textbf{Stellar Mass} \\
 & & & & [\AA] & & [$\log(\text{M}/\text{M}_\odot)$] \\ 

53 & 53.158387 & -27.786760 & 4.774783 & 125.838 $\pm$ 1.429 & 7.852 $\pm$ 2.656 & 8.644 $\pm$ 0.022 \\
375 & 53.156565 & -27.781538 & 3.595535 & 8.458 $\pm$ 2.355 & - & - \\
619 & 53.164854 & -27.779368 & 5.787481 & 277.685 $\pm$ 21.883 & - & 6.388 $\pm$ 0.058 \\
1423 & 53.160882 & -27.801110 & 3.601604 & 18.192 $\pm$ 1.439 & 8.395 $\pm$ 2.144 & 9.065 $\pm$ 0.009 \\
2211 & 53.144195 & -27.785125 & 3.468599 & 101.338 $\pm$ 19.451 & - & 7.901 $\pm$ 0.059 \\
2302 & 53.180459 & -27.770580 & 5.030246 & 106.236 $\pm$ 11.018 & 7.507 $\pm$ 2.361 & 7.191 $\pm$ 0.066 \\
2316 & 53.184491 & -27.780817 & 3.470400 & 90.073 $\pm$ 14.553 & 7.537 $\pm$ 2.938 & 8.572 $\pm$ 0.009 \\
2873 & 53.172484 & -27.764302 & 5.048241 & 83.822 $\pm$ 9.006 & 6.974 $\pm$ 3.532 & - \\
3800 & 53.154222 & -27.796029 & 4.819273 & 245.850$\pm$ 30.364 & 7.297 $\pm$ 3.201 & 6.939 $\pm$ 0.080 \\
4087 & 53.183837 & -27.777307 & 4.641532 & 264.713 $\pm$ 67.949 & - & - \\
4894 & 53.178686 & -27.769167 & 5.618435 & 119.922 $\pm$ 15.419 & 6.666 $\pm$ 3.834 & 6.841 $\pm$ 0.075 \\
5125 & 53.149413 & -27.786889 & 4.516525 & 17.974 $\pm$ 2.563 & - & 6.142 $\pm$ 0.074 \\
5586 & 53.188444 & -27.782993 & 4.050315 & 13.388 $\pm$ 2.847 & - & 6.041 $\pm$ 0.109 \\
6321 & 53.163410 & -27.780357 & 3.758894 & 7.480 $\pm$ 1.362 & - & -  \\
6326 & 53.162342 & -27.777426 & 5.916790 & 453.255 $\pm$ 79.291 & - & - \\
6330 & 53.160611 & -27.778895 & 3.747750 & 87.096 $\pm$ 92.449 & 8.193 $\pm$ 2.318 & -  \\
6333 & 53.157005 & -27.787149 & 5.686581 & 211.488 $\pm$ 32.001 & - & -  \\
6340 & 53.153363 & -27.781371 & 4.508876 & 104.403 $\pm$ 22.129 & - & -  \\
6341 & 53.152609 & -27.783816 & 4.780369 & 23.155 $\pm$ 12.026 & - & - \\
6375 & 53.143162 & -27.787965 & 5.092773 & 5.510 $\pm$ 2.275 & - & - \\
6388 & 53.171560 & -27.765958 & - & - & - & - \\
6473 & 53.167149 & -27.788162 & 2.996250 & 32.764 $\pm$ 34.048 & 6.732 $\pm$ 3.745 & -  \\
6478 & 53.184439 & -27.779954 & 3.748776 & 57.752 $\pm$ 8.704 & - & 9.413 $\pm$ 0.006 \\
6492 & 53.151779 & -27.799840 & 3.979331 & 375.503$\pm$ 88.470 & - & - \\
6510 & 53.154750 & -27.795179 & 3.277048 & 10.232 $\pm$ 3.429 & - & - \\
6862 & 53.185569 & -27.783245 & 3.559977 & 42.603 $\pm$ 9.757 & - & - \\
7203 & 53.141675 & -27.784612 & 3.333151 & 195.109 $\pm$ 39.849 & - & 7.568 $\pm$ 0.013 \\
7287 & 53.150526 & -27.792935 & 4.469047 & 103.470 $\pm$ 11.385 & - & 6.669 $\pm$ 0.052 \\
7685 & 53.161797 & -27.781435 & 5.863154 & 12.533 $\pm$ 2.177 & - & - \\
7755 & 53.150042 & -27.790755 & 3.408508 & 0.523 $\pm$ 0.204 & - & -  \\
7886 & 53.151035 & -27.766202 & 4.924225 & 84.553 $\pm$ 17.120 & - & -  \\
7903 & 53.150166 & -27.783078 & 5.131487 & 18.815 $\pm$ 6.650 & - & - \\
7939 & 53.147156 & -27.793091 & 5.564408 & 15.987 $\pm$ 6.921 & 8.038 $\pm$ 2.142 & 6.940 $\pm$ 0.047 \\
7947 & 53.148288 & -27.785919 & 5.647028 & 10.935 $\pm$ 7.006 & - & - \\
7952 & 53.148764 & -27.763868 & 5.761307 & 76.381 $\pm$ 17.220 & - & 7.096 $\pm$ 0.056 \\
7973 & 53.155866 & -27.793750 & 5.950382 & 72.202 $\pm$ 14.052 & 7.845 $\pm$ 2.367 & 7.651 $\pm$ 0.013 \\
7984 & 53.179287 & -27.773309 & 6.080163 & 75.124 $\pm$ 11.452 & - & 7.694 $\pm$ 0.068 \\
7994 & 53.156396 & -27.800452 & 6.313447 & 107.792 $\pm$ 29.335 & - & 6.690 $\pm$ 0.119 \\
8004 & 53.151784 & -27.800666 & 6.536580 & 18.400 $\pm$ 4.534 & - & - \\
8005 & 53.185732 & -27.774032 & 6.547860 & 450.229 $\pm$ 158.297 & - & -\\
8159 & 53.158386 & -27.779982 & 5.012147 & 128.464 $\pm$ 47.520 & - & - \\
8182 & 53.166273 & -27.791069 & 5.134761 & 3.644 $\pm$ 0.535 & - & - \\
8185 & 53.162846 & -27.792037 & 5.044555 & 4.550 $\pm$ 0.788 & - & - \\
8191 & 53.161589 & -27.782704 & 4.752540 & 3.281 $\pm$ 0.635 & - & - \\
8193 & 53.154870 & -27.784204 & 5.537842 & 7.032 $\pm$ 1.151 & - & - \\
8198 & 53.158826 & -27.783315 & 4.538993 & 5.620 $\pm$ 0.831 & - & - \\
8202 & 53.163787 & -27.781204 & 5.784948 & 2.376 $\pm$ 0.359 & - & - \\
8203 & 53.166863 & -27.780926 & 5.782297 & 9.375 $\pm$ 1.243 & - & - \\
8207 & 53.171260 & -27.785981 & 4.267632 & 23.000 $\pm$ 5.825 & - & - \\
8246 & 53.165294 & -27.780259 & 3.001163 & 5.007 $\pm$ 1.405 & - & - \\
8256 & 53.165169 & -27.787870 & 4.750995 & 98.177 $\pm$ 21.279 & - & 7.065 $\pm$ 0.075 \\
8264 & 53.158952 & -27.792315 & 5.899146 & 4.097 $\pm$ 0.700 & - & -  \\
8283 & 53.159455 & -27.792426 & 3.953544 & 5.927 $\pm$ 1.154 & - & - \\
8284 & 53.168184 & -27.790814 & 3.110677 & 0.512 $\pm$ 66.726 & - & - \\
8288 & 53.159517 & -27.781704 & 3.659778 & 11.624 $\pm$ 2.252 & - & - \\
8290 & 53.153692 & -27.790357 & 3.601366 & 91.064 $\pm$ 10.205 & - & 8.093 $\pm$ 0.018 \\
8296 & 53.167744 & -27.789259 & 4.554394 & 5.148 $\pm$ 0.830 & - & - \\
8307 & 53.160208 & -27.786649 & 4.140301 & 7.804 $\pm$ 1.408 & - & - \\
8320 & 53.164416 & -27.790148 & 3.187705 & 28.126 $\pm$ 7.999 & - & - \\
8328 & 53.160899 & -27.787982 & 4.715065 & 20.177 $\pm$ 2.963 & - & 6.911 $\pm$ 0.168 \\
8339 & 53.169816 & -27.788370 & 3.194243 & 38.430 $\pm$ 8.302 & - & - \\
8342 & 53.171449 & -27.787623 & 3.587813 & 76.893 $\pm$ 27.512 & -  & - \\
8431 & 53.164918 & -27.788481 & 5.894479 & 12.776 $\pm$ 1.682 & 7.469 $\pm$ 2.459 & - \\
8450 & 53.162971 & -27.787537 & 3.746559 & 2.334 $\pm$ 0.642 & - & - \\
29012 & 53.190746 & -27.778036 & 3.007549 & 10.143 $\pm$ 0.539 & 8.031 $\pm$ 2.485 & 8.517 $\pm$ 0.006 \\
34901 & 53.164803 & -27.749762 & 4.773070 & 161.951$\pm$ 5.318 & 7.222 $\pm$ 3.280 & 8.197 $\pm$ 0.068 \\
35777 & 53.161060 & -27.745220 & 3.698426 & 392.629 $\pm$ 39.469 & - & 7.296 $\pm$ 0.031 \\
36231 & 53.172543 & -27.743000 & 3.920429 & 101.748 $\pm$ 4.782 & - & 8.701 $\pm$ 0.012 \\
136039190 & 53.128847 & -27.767184 & 4.941978 & 505.766 $\pm$ 30.652 & - & 6.832 $\pm$ 0.075 \\
137026047 & 53.124333 & -27.763214 & 3.599473 & 53.185 $\pm$ 9.443 & 7.376 $\pm$ 3.122 & 6.091 $\pm$ 0.053 \\
140006025 & 53.129543 & -27.758040 & 3.490443 & 92.812 $\pm$ 17.289 & - & - \\
140012041 & 53.134896 & -27.747224 & 3.792470 & 134.856 $\pm$ 22.695 & - & - \\
140018054 & 53.132585 & -27.752442 & 3.972120 & 262.450 $\pm$ 20.982 & 7.353 $\pm$ 3.156 & 9.308 $\pm$ 0.008 \\
147002007 & 53.155239 & -27.746955 & 2.939298 & 73.172 $\pm$ 17.085 & - & - \\
147049145 & 53.150317 & -27.738307 & 4.490233 & - & - & 6.792 $\pm$ 0.092 \\
148032095 & 53.170672 & -27.744212 & 4.601113 & - & - & 7.876 $\pm$ 0.012 \\
148039109 & 53.158098 & -27.736889 & 4.867925 & 41.053 $\pm$ 1.611 & - & 8.976 $\pm$ 0.001 \\
148045118 & 53.157562 & -27.744125 & 5.080155 & 100.657 $\pm$ 2.826 & - &  8.302 $\pm$ 0.028 \\
149058240 & 53.179249 & -27.741977 & 6.034911 & -  & - & -  \\
150007031 & 53.181409 & -27.746872 & 3.285502 & 367.176 $\pm$ 68.793 & -  & 7.181 $\pm$ 0.030 \\
151008047 & 53.183251 & -27.761966 & 3.195008 & -  & - & 9.075 $\pm$ 0.002 \\
151019099 & 53.184700 & -27.761023 & 3.602653 & -  & 8.048 $\pm$ 2.360 & 8.945$\pm$ 0.008 \\
151025112 & 53.194009 & -27.762054 & 3.885642 & 114.036 $\pm$ 12.717 & 7.572 $\pm$ 2.908 & 6.695 $\pm$ 0.043 \\
151044143 & 53.193991 & -27.768969 & 4.864627 & - & 8.096 $\pm$ 2.273 & 9.205 $\pm$ 0.004 \\
152009065 & 53.200721 & -27.772766 & 4.027376 & 13.722 $\pm$ 2.032 & 7.866 $\pm$ 2.605 & - \\
152025104 & 53.190050 & -27.777804 & 3.664041 & 615.030 $\pm$ 100.438 & - & 6.668 $\pm$ 0.048 \\
152032118 & 53.196305 & -27.776388 & 3.944359 & 472.733 $\pm$ 25.320 & 7.332 $\pm$ 3.170 & 7.262 $\pm$ 0.017 \\
154014100 & 53.179764 & -27.760319 & 3.773295 & 1089.787 $\pm$ 95.765 & 7.495 $\pm$ 3.007 & 6.936 $\pm$ 0.018 \\

\end{longtblr}

    %\input{table1}

%\subsection{Ly$\alpha$ EWs}

    %\input{table2}

%%%%%%%%%%%%%%%%%%%%%%%%%%%%%%%%%%%%%%%%%%%%%%%%%%

% Don't change these lines
\bsp	% typesetting comment
\label{lastpage}
\end{document}